\documentclass[a4paper,fleqn,usenatbib,useAMS]{mnras}

\usepackage{graphicx}	
\usepackage{amsmath}	
\usepackage{amssymb}	
\usepackage{multicol}        
\usepackage{bm}		
\usepackage{pdflscape}	

\usepackage[T1]{fontenc}
\usepackage{ae,aecompl}

\usepackage{newtxtext,newtxmath}

\title[Strong magnetic field in delta-type sunspot]
{Formation of super-strong horizontal magnetic field in delta-type sunspot in radiation magnetohydrodynamic simulations}

\author[H. Hotta and S. Toriumi]{  
  H. Hotta$^{1}$\thanks{Contact e-mail: \href{mailto:hotta@chiba-u.jp}{hotta@chiba-u.jp}} and
  S. Toriumi$^{2}$\thanks{Contact e-mail: \href{mailto:toriumi.shin@jaxa.jp}{toriumi.shin@jaxa.jp}}
\\
$^{1}$Department of Physics, Graduate School of Science, Chiba University, 1-33 Yayoi-cho, Inage-ku, Chiba 263-8522, Japan \\
$^{2}$Institute of Space and Astronautical Science (ISAS)/Japan Aerospace Exploration Agency (JAXA), 3-1-1 Yoshinodai, \\
~Chuo-ku, Sagamihara, Kanagawa 252-5210, Japan
}

\date{}

\pubyear{}

\begin{document}
\label{firstpage}
\pagerange{\pageref{firstpage}--\pageref{lastpage}}
\maketitle

\begin{abstract}
  We perform a series of radiative magnetohydrodynamic simulations to understand the amplification mechanism of the exceptionally strong horizontal magnetic field in delta-type sunspots. In the simulations, we succeed in reproducing the delta-type sunspot and resulting strong magnetic field exceeding 6000 G in a light bridge between the positive and negative polarities. Our conclusions in this study are summarized as follows:
  1. The essential amplification mechanism of the strong horizontal magnetic field is the shear motion caused by the rotation of two spots.
  2. The strong horizontal magnetic field remains the force-free state. 
  3. The peak strength of the magnetic fields does not depend on the spatial resolution, top boundary condition, or Alfven speed limit.
  The origin of the rotating motion is rooted in the deep convection zone.
  Therefore, the magnetic field in the delta-spot light bridge can be amplified to the superequipartition values in the photosphere.
\end{abstract}

\begin{keywords}
  Sun: magnetic fields -- Sun: photosphere -- Sun: sunspots
\end{keywords}


\section{Introduction}
Sunspots are the most prominent feature on the solar surface. Because of the strong magnetic field in the sunspots, the convection energy transport is significantly suppressed, and the sunspot area is darkened.
Typically, the strongest magnetic field is observed at the centre of the spot, i.e. the darkest region (umbra) \citep{1996A&A...316..229K,2003A&ARv..11..153S}. The maximum magnetic field strength is generally 2500 G, and horizontal magnetic fields reach 1000 G around the penumbra/quiet-sun boundary of sunspots. \cite{2006SoPh..239...41L} reports a 6100~G magnetic field in the umbra as the strongest field from the statistical data collected between 1917 and 2004.
\par
Several exceptions, i.e. the strong magnetic fields outside the umbra, are also reported. \cite{1991SoPh..136..133T} discovers a 4300 G horizontal magnetic field in a delta-type spot using Big Bear Solar Observatory data in 1971. The strong magnetic field locates at a light bridge between the positive and negative spots. \cite{1993SoPh..144...37Z} report several observations and also show a strong (>3500~G) horizontal magnetic field around the polarity inversion line (PIL). More recently, \cite{2018ApJ...852L..16O} analyse the Hinode SOT/SP data of NOAA  11967 and through the Milne--Eddington inversion technique recover a field strength of 6250 G at a light bridge along the PIL. Later, \cite{2020arXiv200312078C} apply the stratified inversion method that takes into account the point spread function of SOT to the same data set as \cite{2018ApJ...852L..16O}, finding that the strongest field strength is 8200 G at the $\tau=1$ layer, where $\tau$ is the optical depth.
\par
At the solar surface, the typical density is $\rho=2\times10^{-7}~\mathrm{g~cm^{-3}}$, the typical convection velocity is $v_\mathrm{c}=4~\mathrm{km~s^{-1}}$, and the typical gas pressure is $p=7.6\times10^4~\mathrm{dyn~cm^{-2}}$. 
These lead to the equipartition magnetic field strengths against the kinetic $B_\mathrm{eq(kin)}$ and internal $B_\mathrm{eq(int)}$ energies of
\begin{align}
  B_\mathrm{eq(kin)}=&\sqrt{4\pi \rho v_\mathrm{c}^2}\sim 600~\mathrm{G}, \\
  B_\mathrm{eq(int)}=&\sqrt{8\pi p}\sim 1400~\mathrm{G},
\end{align}
respectively. Thus, the magnetic field strength of more than 6000 G is significantly superequipartition. 
Flow and gas pressure originating in the photosphere are not sufficient to amplify and maintain such a strong magnetic field. 
We use the average velocity and gas pressure for this discussion. These can be larger locally and amplify the magnetic field more on a small scale. For example, the small-scale dynamo calculation shows a field of 2500 G at maximum \citep{2014ApJ...789..132R}. The value is still not enough to explain the observed 6000 G. Also, the strong magnetic field found in the observations has a larger spatial scale than the convection scale, and we need some coherent amplification mechanism(s).
\cite{2018ApJ...852L..16O} suggest that the compression by the Evershed flow from one spot to the other produces the observed strong magnetic field. However, because the creation of such a strong field may not only be caused by the surface mechanisms but is probably linked to the dynamics in the deep layers, a detailed radiation magnetohydrodynamic simulation that can address the spot generation from the deep convection zone is needed.
\par
The flux emergence and sunspot formation involve the radiation, the convection, the ionization, and the stratification. There have been several calculations that include all these processes \citep{2010ApJ...720..233C,2012ApJ...753L..13S,2014ApJ...785...90R,2017ApJ...846..149C}. These studies basically model regular sunspots rather than the delta-type sunspot, and their bottom boundary is located in a relatively shallow layer (<30 Mm). In addition, magnetic flux is often kinematically injected from the bottom boundary.
\par
Recently, we succeed in reproducing delta-type spots in a deep domain calculation \citep[][hereafter TH19]{2019ApJ...886L..21T}. A large-scale flow in the deep region causes the collision of two spots with opposite polarities, and the delta-type spot is created in the photosphere \citep[see also][]{2016PhRvL.116j1101C}.
We have already found a strong magnetic field ($\sim$4000 G) in TH19.
In this study, we adopt the simulation setup similar to TH19 but increase the magnetic flux to explore the possibility of a stronger magnetic field in the delta-type spot. Then, the amplification mechanism of such a field is studied. To verify the validity of the amplified magnetic field in our simulation, we change the several numerical settings such as the resolution, top boundary condition, and Alfven speed limit.
\par
The rest of the paper is structured as follows. Section \ref{Model} describes the setting of the numerical simulation. Section \ref{Result} shows the calculation results of the formation of the strong horizontal magnetic field and its amplification mechanism. In Section \ref{Summary}, we summarize our results.

\section{Model}
\label{Model}
\begin{figure}
  \centering
  \includegraphics[width=0.5\textwidth]{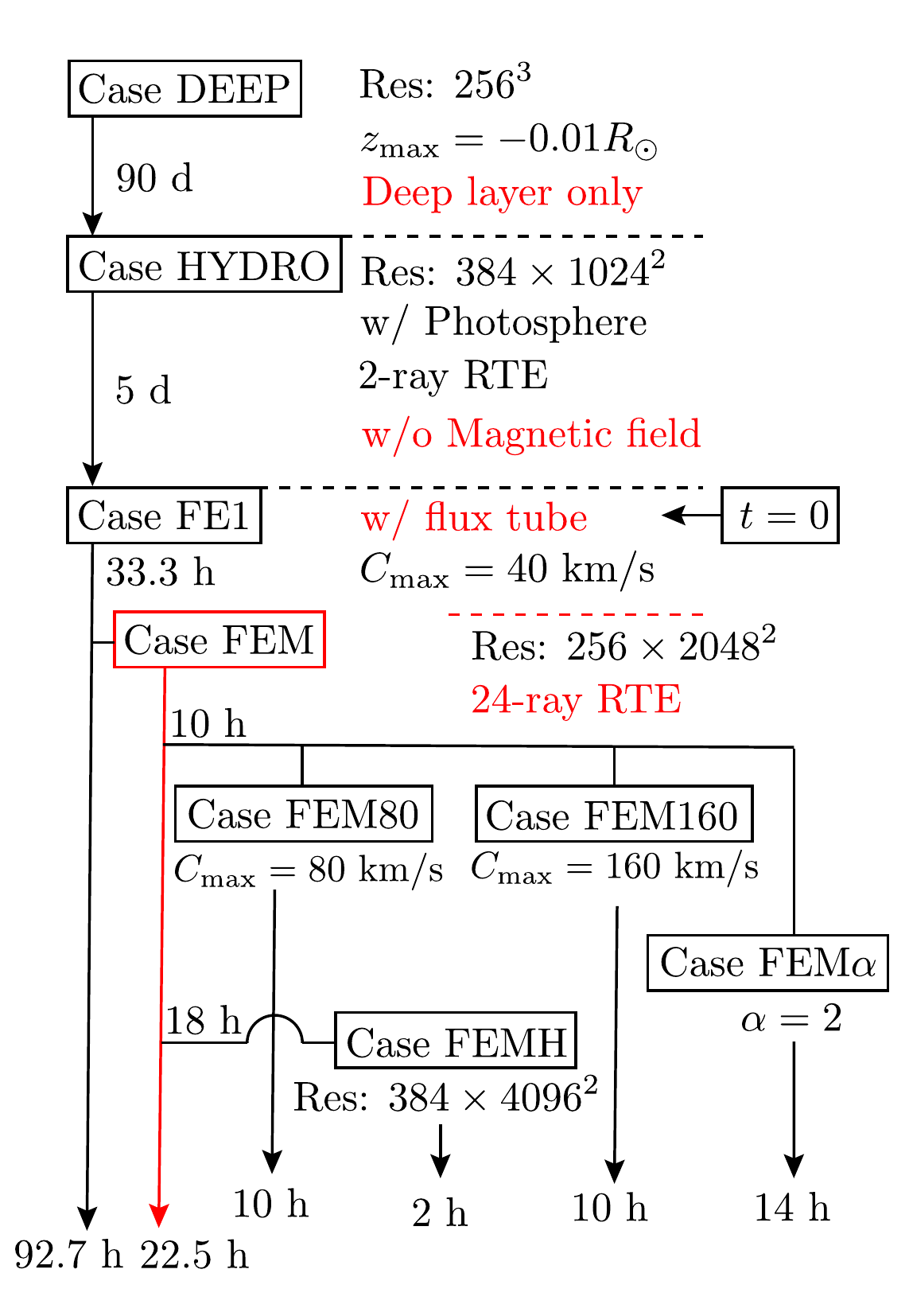}
  \caption{Calculation procedure.}
  \label{chart}
\end{figure}

\begin{table*}
  \centering
  \caption{Overview of numerical simulations}
  \begin{tabular}{lccccccc}    
    \hline \hline
     Case & No. of grids & period & $z_\mathrm{max}$ & magnetic field & Radiation transfer & $C_\mathrm{max}$ & $\alpha$ \\
     \hline
     DEEP      & $256\times256\times256$   & 90 d   &$-0.01R_\odot$ & no & N/A & N/A  & N/A \\
     HYDRO     & $384\times1024\times1024$ & 5 d    & 700 km & no  & one ray    & N/A & N/A \\
     FE1       & $384\times1024\times1024$ & 33.3 h & 700 km & yes & one ray    & 40  $\mathrm{km~s^{-1}}$ & 1 \\
     FEM       & $256\times2048\times2048$ & 22.5 h & 700 km & yes & multi rays & 40  $\mathrm{km~s^{-1}}$ & 1 \\
     FEM80     & $256\times2048\times2048$ & 10 h   & 700 km & yes & multi rays & 80  $\mathrm{km~s^{-1}}$ & 1 \\
     FEM160    & $256\times2048\times2048$ & 10 h   & 700 km & yes & multi rays & 160 $\mathrm{km~s^{-1}}$ & 1 \\
     FEM$\alpha$& $256\times2048\times2048$ & 10 h  & 700 km & yes & multi rays & 40  $\mathrm{km~s^{-1}}$ & 2 \\
     FEMH      & $384\times4096\times4096$ & 2 h    & 700 km & yes & multi rays & 40  $\mathrm{km~s^{-1}}$ & 1 \\
    \hline
  \end{tabular}
\end{table*}

\begin{figure}
  \centering
  \includegraphics[width=0.5\textwidth]{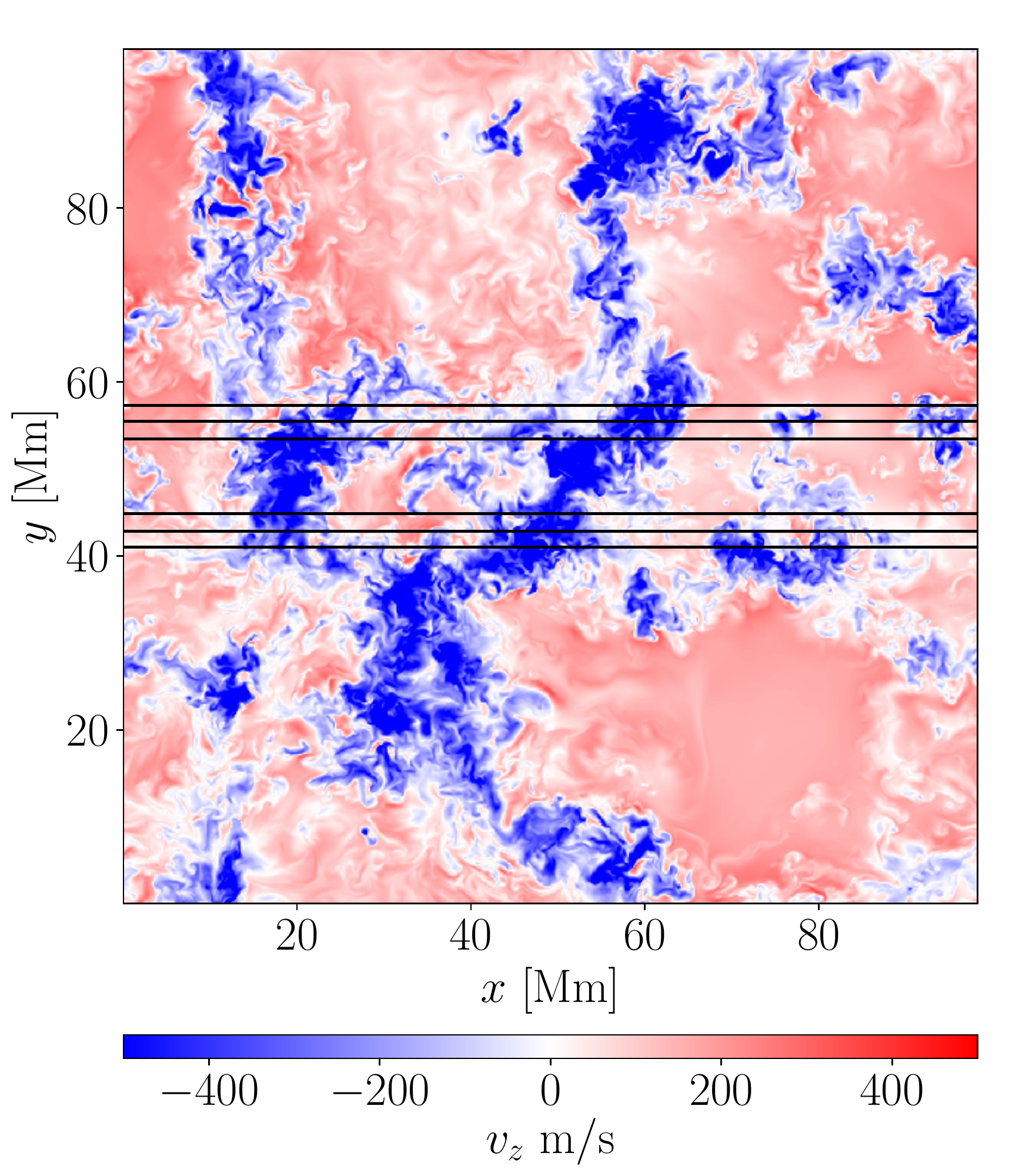}
  \caption{The initial location of the flux tube in Case FE1. Quantities at $z=-22~\mathrm{Mm}$, where the initial flux tube is inserted. The colour contour shows the vertical velocity $v_z$. The black contour lines show $B_y$, which correspond to 2500, 5000, and 7500~G.}
  \label{initial}
\end{figure}

We solve the three-dimensional magnetohydrodynamic equation with the radiation using the R2D2 code \citep[][hereafter HI20]{2019SciA....eaau2307,2020MNRAS.494.2523H} in the Cartesian geometry ($x$,$y$,$z$), where $x$ and $y$ are the horizontal directions and $z$ is the vertival direction. $z=0$ indicates the photosphere. To relax the constraint on the time step by the fast speed of sound in the deep convection zone, we adopt the RSST \citep{2012A&A...539A..30H,2015ApJ...798...51H}. The overall settings are the same as those in HI20. In this study, we only adopt the artificial viscosity and thus there is no explicit parameter for the diffusion.\par
We also use the Alfven speed limit in the low-$\beta$ region above the photosphere in order to relax a severe constraint by the CFL condition with the fast Alfven wave there. \cite{2009ApJ...691..640R} suggests the method by suppressing the Lorentz force. In most cases, we use the Alfven speed limit $C_\mathrm{max}$ of 40~$\mathrm{km~s^{-1}}$. For the purpose of this study, i.e. finding the strength of the magnetic field in the photosphere, it is important to discuss the validity of this method.
Thus, we vary $C_\mathrm{max}$ to 80 and 160$~\mathrm{km~s^{-1}}$ in two cases.
\par
According to \cite{2012ApJ...750...62R}, the appearance of penumbral structures is significantly affected by the choice of the top boundary condition. We here follow \cite{2012ApJ...750...62R}'s definition of the parameter $\alpha$, which controls the top boundary condition. 
The parameter $\alpha$ is multiplied to the horizontal components of the potential magnetic field, $B_x$ and $B_y$, to control the inclination of the field, and the vertical distribution of the field is modified in such a way that it satisfies the divergence free condition (see eq. (B3) of \cite{2012ApJ...750...62R} for the detailed definition of $\alpha$). For most of the cases that include the photosphere, we adopt $\alpha=1$, i.e. the top boundary is the potential magnetic field. In one exceptional case, where we still have the photosphere, the top boundary condition is chosen to be $\alpha=2$, making the magnetic field more horizontal than the potential field. In one case, we choose $\alpha=2$, in which the magnetic field at the top boundary is more horizontal than the potential magnetic field.
\par
The calculation domain extends 98.304 Mm in the horizontal direction and 202.537 Mm in the vertical direction except for Case DEEP. For all the cases, the bottom boundary is at the base of the convection zone ($z=-0.29R_\odot$, where $R_\odot=696 \mathrm{~Mm}$).
\subsection{Calculation procedure}
Our calculation procedure is shown in Fig. \ref{chart}.
We start our calculations from the low-resolution deep convection zone run that lacks the photosphere with the top boundary at 0.99$R_\odot$, resolved by the number of grid points of $256^3$ (Case DEEP). The computation speed is accelerated by omitting the photosphere because we can circumvent the calculation of small-scale, short-lived, granular convection.
Case DEEP continues for 90 days. The typical convection time scale in the deep convection zone is 30 days and 90 days corresponds to three turn over times. However, \cite{2019SciA....eaau2307} show that calculations for 75 days and 360 days yield almost the same stratification. Thus we conclude that 90 days is enough for relaxation.
\par
The calculation domain is extended to include the photosphere (Case HYDRO). From here, the top boundary is at $700~\mathrm{km}$ above the average $\tau=1$ surface. The resolution is upgraded to $384\times1024^2$. The horizontal grid spacing is 96 km, which is acceptable for resolving the photosphere. We use non-uniform grid spacing in the vertical direction, which is 48 km and 1.5 Mm at the photosphere and the base of the convection zone, respectively. This grid spacing 
of 1.5 Mm is acceptable for resolving the deep convection zone, where the pressure scale height is 60 Mm. Case HYDRO continues for five days. Then, statistically steady convection covering the whole convection zone is prepared. This approach is justified because the near-surface layer does not have a significant influence on the deep convection \citep{2019SciA....eaau2307}.
\par
From Case FE1, we start the flux emergence simulations. The force-free twisted flux tube is inserted at a depth of 22 Mm. We use the Bessel function for the force-free flux tube (see HI20 for the details). The axial magnetic field at the centre of the flux tube and the radius of the flux tube are $10^4~\mathrm{G}$ and 10~Mm, respectively. The total magnetic flux is $1.35\times10^{22}~\mathrm{Mx}$, which is twice larger than that in TH19 to have stronger magnetic field in the photosphere.
We define the beginning of Case FE1 as $t=0$. In this case, we only solve two rays for the radiation transfer; thus, only upward and downward energy transfers are allowed. Case FE1 continues for 92.7~h. The initial location of the flux tube is shown in Fig. \ref{initial}. The coherent large-scale downflow at the centre of the calculation domain pins down the middle of the flux tube, whereas the two ends emerge to the photosphere. As a result, the sunspots of opposite polarities collide with each other to generate a delta-type spot in a self-consistent manner \citep[the multi-buoyant segment scenario: see][and TH19 for the details]{2014SoPh..289.3351T}.
\par
After the initial sunspots are created at $t=33.3~\mathrm{h}$, we upgrade the calculation to Case FEM, in which the number of grids is $256\times2048^2$. The horizontal grid spacing is 48 km. We keep the 48~km vertical grid spacing in Case FEM in the photosphere, but the grid spacing at the base of the convection zone is made coarse by using a spacing of 3 Mm because the details of the convection in the deep layer do not have a significant influence on the sunspot evolution in the photosphere.  From here, we use the multi-ray radiation transfer. We solve 24 rays (see Appendix \ref{multi_ray} for the details of the calculation scheme). 
The one-ray radiation transfer prohibits the horizontal radiation energy transfer, and the convection pattern tends to show unrealistic small-scale features in high-resolution calculations.
In addition, we adopt the potential magnetic field at the top boundary ($\alpha=1$). In this study, we mainly analyse the Case FEM.
\par
By using the data at $t=43.3~\mathrm{h}$ of the Case FEM, we restart Cases FEM80, FEM160, and FEM$\alpha$. In Cases FEM80 and FEM160, we change the Alfven speed limit to $C_\mathrm{max}=80$ and $160~\mathrm{km~s^{-1}}$, respectively. In Case FEM$\alpha$, we adopt $\alpha=2$, in which the magnetic field at the top boundary is more horizontal than the potential magnetic field. Cases FEM80 and FEM160 continue for 10 h, whereas Case FEM$\alpha$ continues for 14 h. In these cases, the other settings are the same as those of Case FEM.
\par
\begin{figure}
  \centering
  \includegraphics[width=0.5\textwidth]{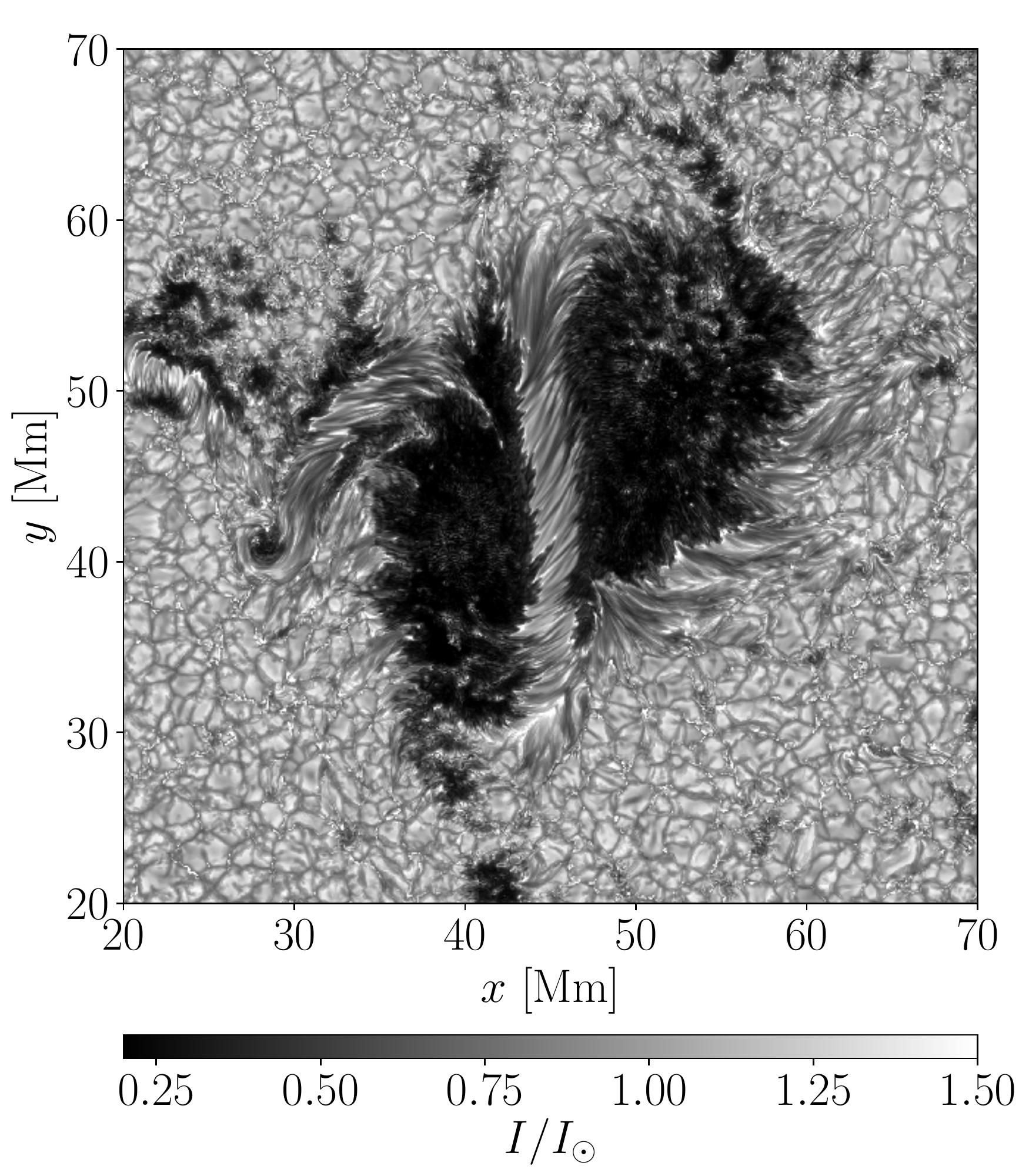}
  \caption{The emergent intensity at $t=53.3~\mathrm{h}$ in the highest resolution case (Case FEMH). A movie for Case FEMH is available online.}
  \label{display}
\end{figure}

The data at $t=51.3~\mathrm{h}$ of Case FEM are upgraded to a higher resolution and restarted as Case FEMH. The number of grid points in Case FEMH is $384\times4096^2$. The grid spacing in the photosphere is 24 km in all three directions. We still use the non-uniform vertical grid spacing, which is 3 Mm at the base of the convection zone. Case FEMH continues for 2 h. Fig. \ref{display} shows the emergent intensity in Case FEMH.
\par
The change of resolution implies the difference in the numerical diffusivity. With a higher resolution, we can capture small-scale features. Since the difference of the resolution mainly influences the small scales, which have much shorter time scales, the relaxation occurs in much smaller time scales than the typical convection time scale, which is several minutes in the photosphere.

\section{Result}
\label{Result}
\subsection{Overall evolution}
\begin{figure}
  \centering
  \includegraphics[width=0.5\textwidth]{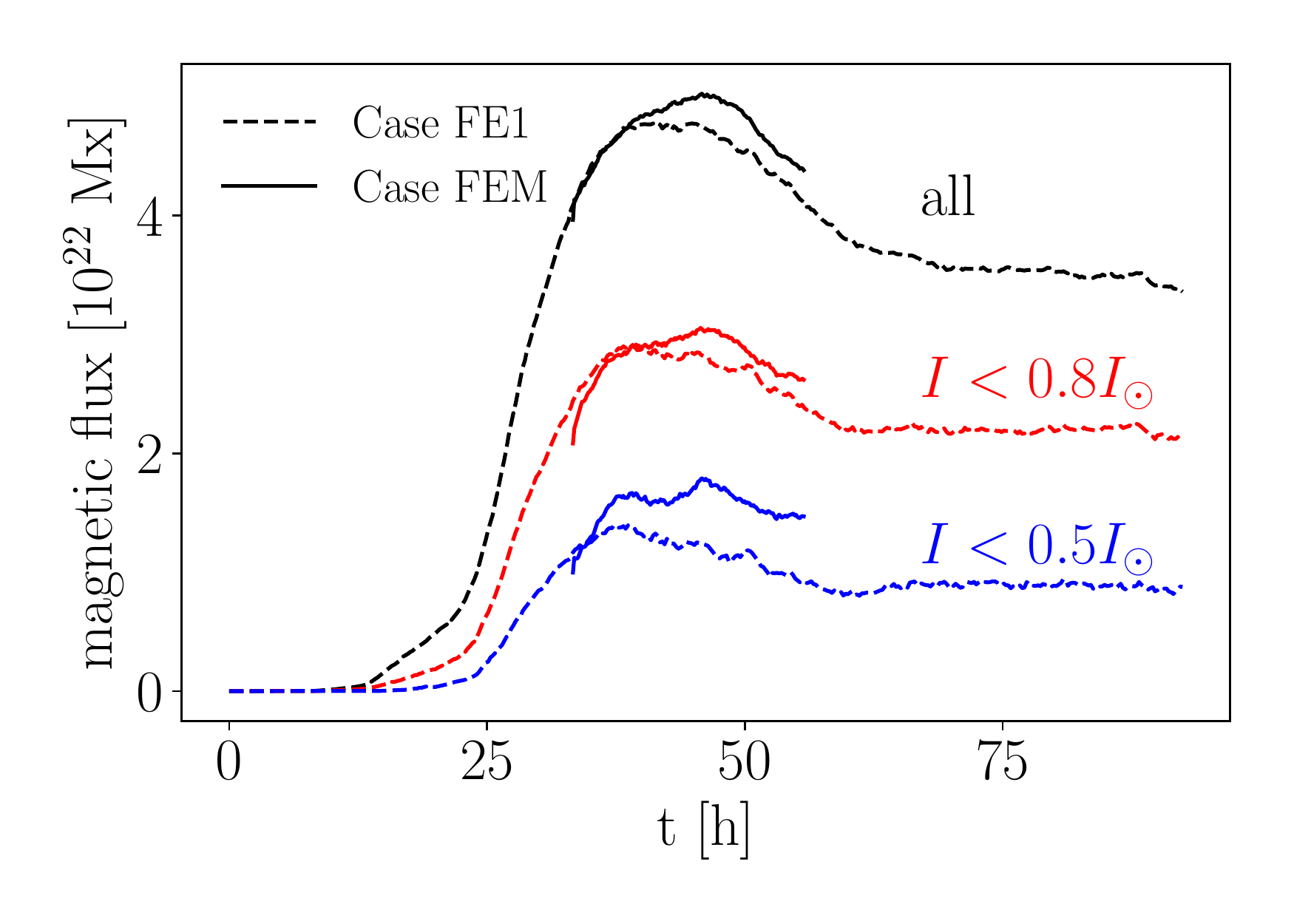}
  \caption{Evolution of the unsigned magnetic flux on the $\tau=1$ surface. The solid and dashed lines show the results of Cases FEM and FE1, respectively. The black line indicates the magnetic flux over the whole computational domain. The red and blue lines are the magnetic flux in regions with intensities less than 80\% and 50\% of the average photospheric intensity $I_\odot$, respectively.}
  \label{magnetic_flux}
\end{figure}

In this subsection, the overall evolution of the flux tube and the generated sunspots are explained.
Fig. \ref{magnetic_flux} shows the evolution of the unsigned magnetic flux at the $\tau=1$ surface. 
The solid and the dashed lines indicate the results from Cases FEM and FE1, respectively.
The black lines are the unsigned magnetic flux in the whole computational domain. The red and blue lines show the magnetic flux in the regions with the intensities less than 80\% and 50\% of the averaged photospheric intensity $I_\odot$, respectively, which roughly correspond to the penumbral and umbral regions. The maximum unsigned magnetic flux is almost $5\times 10^{22}~\mathrm{Mx}$, which is well within the solar range \citep[$\lesssim 2\times10^{23}$:][]{2017ApJ...834...56T}, indicating that our setup of the magnetic flux is applicable to the Sun.
 We note that the magnetic flux reaches the steady state after $t=60~\mathrm{h}$, i.e. the magnetic flux does not decrease. This plateau indicates that the amount of photospheric flux in the delta-spot simulations does not decrease as fast as in the other flux emergence simulations (see HI20). However, we do not discuss the spot decay in detail in the present study. The calculation duration of Case FEM is shorter than that of Case FE1 but covers the period of the peak unsigned magnetic flux ($t=45~\mathrm{h}$). The overall evolution of the magnetic flux is similar between Cases FE1 and FEM, but we see a distinct difference, especially in the blue lines ($I<0.5I_\odot$), corresponding to the umbra. This indicates that the resolution and/or the radiation transfer treatment can affect the umbral evolution.
\par
\begin{figure}
  \centering
  \includegraphics[width=0.5\textwidth]{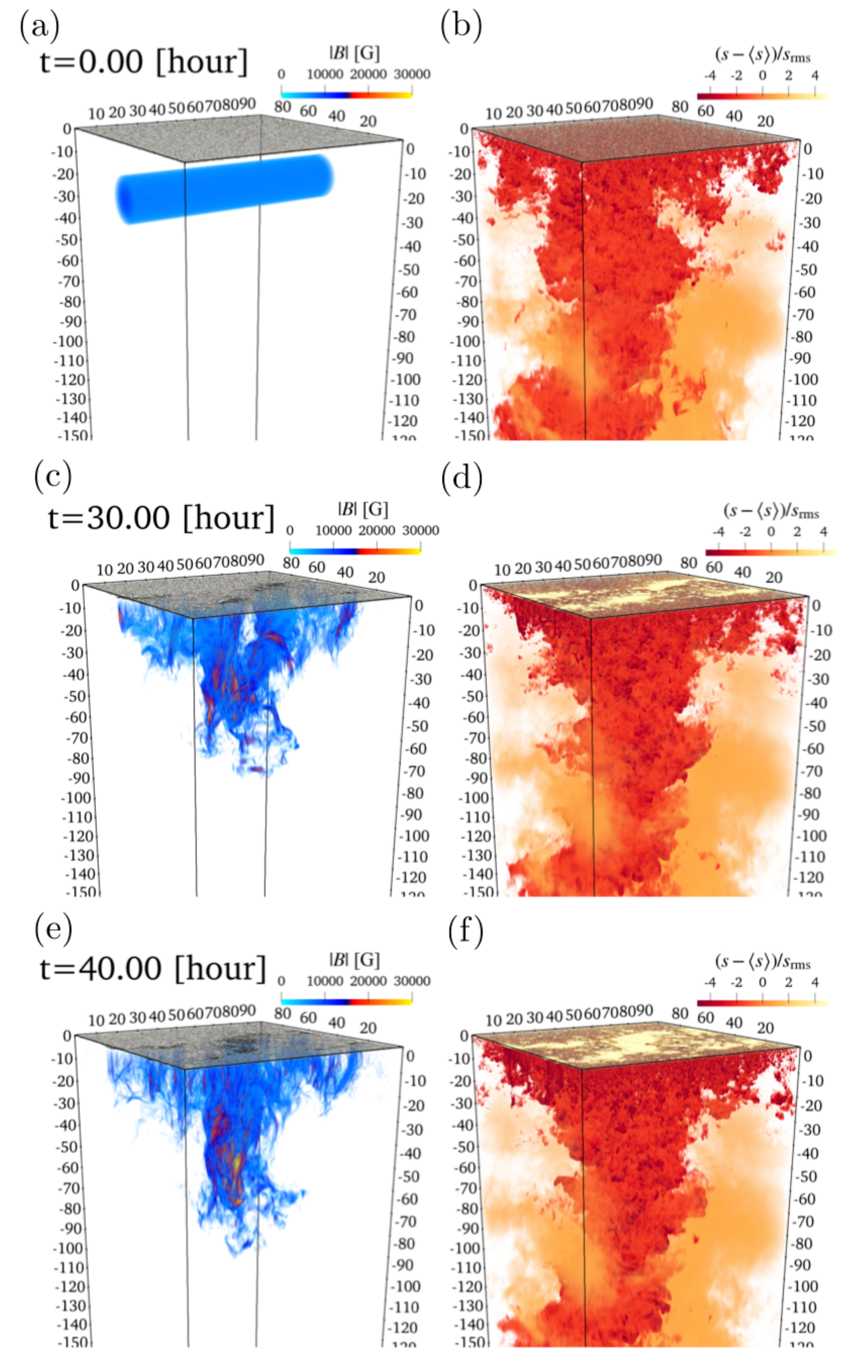}
  \caption{3D volume rendering of the magnetic field strength (panels a, c, and e) and the normalized entropy ($(s-\langle s\rangle)/s_\mathrm{rms}$: panels b, d, and f). The grey surface around the top boundary shows the emergent intensity at the $\tau=1$ surface. A movie for Case FE1 is available online.}
  \label{3d}
\end{figure}

Fig. \ref{3d} shows the evolution of the three-dimensional structure of the flux tube and the normalized entropy in Case FE1. The left panels show the magnetic field strength $|B|$, and the right panels are the normalized entropy $(s-\langle s\rangle)/s_\mathrm{rms}$, where $s$ is the specific entropy. $\langle s\rangle$ and $s_\mathrm{rms}$ are the horizontally averaged and root-mean-square (rms) entropy for Case HYDRO, respectively. The results show a large-scale coherent downflow in the centre of the calculation domain that extends to the deeper layer ($z<-150~\mathrm{Mm}$, red clump at the center of the calculation domain extending from the photosphere to the deep layer, see also the movie). Because the convection time scale in the deeper layer is long ($\sim30~\mathrm{d}$), the overall structure of this downflow does not change during the calculation period of Case FE1 (Figs. \ref{3d}b, d, and f). This long-lived coherent downflow is the essential origin of the delta-type spot in this study. Upflow beside the coherent downflow causes  the rising of the flux tube and the formation of the sunspot in the photosphere. The coherent downflow in the deep region causes the convergent flow between the two sunspots, and they approach each other (Fig. \ref{3d}c). Then, the two sunspots collide to form a delta-type sunspot.
We note that the force-free state is broken as the flux tube starts to rise. The density decreases in the flux tube and the gas pressure gradient plays a significant role in maintaining the force balance of the flux tube (see also HI20)
\par
\begin{figure}
  \centering
  \includegraphics[width=0.5\textwidth]{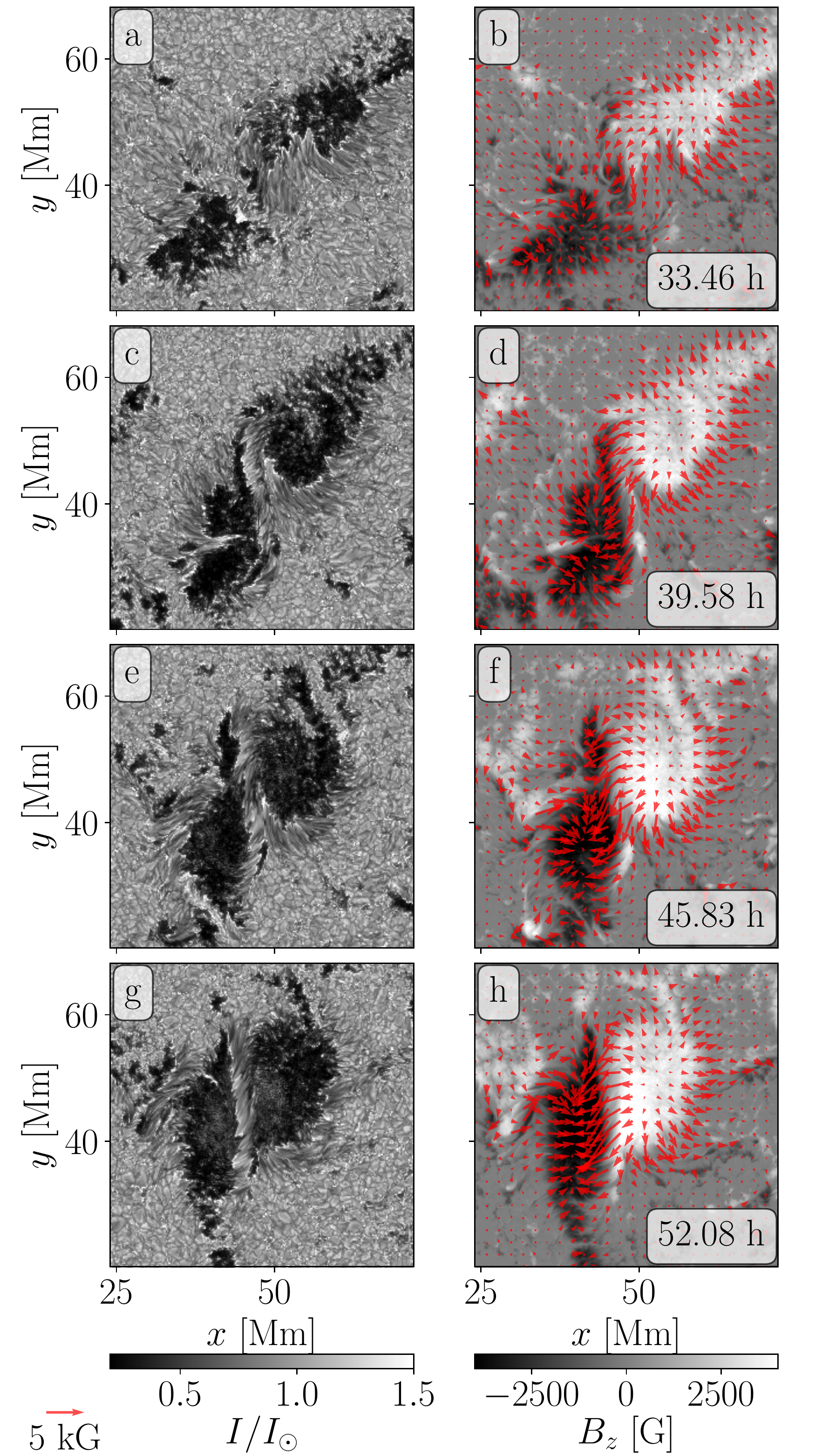}
  \caption{The emergent intensity (panels a, c, e, and g) and the magnetic fields at $\tau=0.01$ (panels b, d, f, and h) in Case FEM. The greyscale and the red arrows show the vertical and horizontal magnetic fields, respectively. A movie for Case FEM is available online.}
  \label{overall}
\end{figure}

Fig. \ref{overall} shows the emergent intensity and the magnetic field on the $\tau=0.01$ surface in Case FEM. When the two sunspots are separated from each other, the sunspot magnetic field is mostly vertical, and the horizontal magnetic field strength is only a few thousand Gauss at maximum (Figs. \ref{overall}a and b). As the two sunspots approach, penumbra-like structures are constructed between them (Fig. \ref{overall}c). The horizontal magnetic field perpendicular to the PIL becomes intensified (Fig. \ref{overall}d). At $t=52~\mathrm{h}$, the two sunspots reach each other, and the horizontal magnetic field exceeds 6000 G.

\par
\begin{figure}
  \centering
  \includegraphics[width=0.5\textwidth]{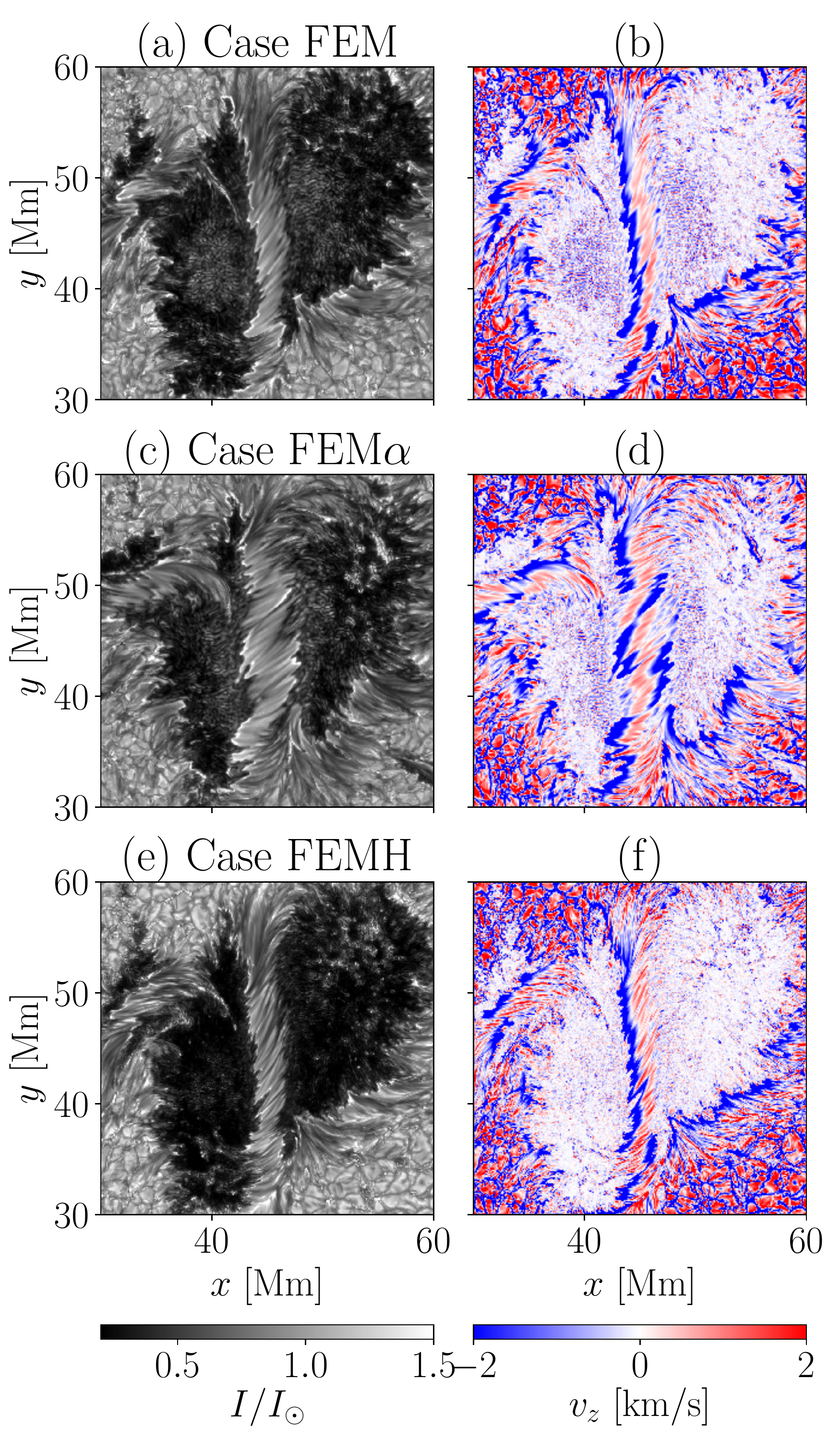}
  \caption{The emergent intensity (panels a, c, and e) and the vertical velocity (panels b, d, and f) in Cases FEM (panels a and b), FEM$\alpha$ (panels c and d), and FEMH (panels e and f). All the data are at $t=53.3~\mathrm{h}$. The vertical velocity is measured on the $\tau=1$ surface. A movie for Case FEM$\alpha$ is available online.}
  \label{comp_reso}
\end{figure}

Fig. \ref{comp_reso} presents a comparison between Cases FEM, FEM$\alpha$, and FEMH. Figs. \ref{comp_reso}a, c, and e indicate the emergent intensity; and b, d, and f depict the vertical velocity on the $\tau=1$ surface. All panels show the quantities at $t=53.3~\mathrm{h}$. While the photospheric surface is filled with fine-scale structures in Case FEMH (panels e and f), the overall structure, such as the width of the light bridge between the two sunspots, is similar to that of the lower-resolution run (Case FEM: panels a and b). The typical width of the filamentary structure across the PIL is 1 Mm for both Cases FEM and FEMH. One notable difference between Cases FEM and FEMH is the intensity in the umbra. Case FEMH (Fig. \ref{comp_reso}e) shows lower intensity in the umbra than that in Case FEM (Fig. \ref{comp_reso}a). \cite{2012ApJ...750...62R} suggests that the numerical diffusivity causes a mass diffusion and the resulting increase of the umbral dot. Case FEMH has less numerical diffusivity than Case FEM, which probably is the cause of the reduced umbral intensity in Case FEMH. This result implies that the actual Sun would show smaller umbral intensity than the simulations shown in this paper since the actual Sun has much smaller diffusivity.
The vertical velocity structures in Cases FEM and FEMH are almost identical. Along the light bridge, one may find the region of weak upflows at the edges of the sunspot on the right. The corresponding region of the left sunspot is filled with downflows.
Some fluctuations may be found in this flow structure, but the upflow--downflow relation does not change along the light bridge. 
\par
Case FEM$\alpha$ (Figs. \ref{comp_reso}c and d) exhibits significantly different features from Cases FEM and FEMH. Specifically, the width of the light bridge in Cases FEM and FEMH is 2--3 Mm, which is larger than 5 Mm in Case FEM$\alpha$. In response to this change, the area of the umbrae is decreased. \cite{2012ApJ...750...62R} reveals that the increase of $\alpha$ extends the length of the penumbra, but the umbral area is not influenced significantly in his comparison \citep[see Figs. 2 and 3 of][]{2012ApJ...750...62R}. We expect that the difference in topology, i.e. whether the sunspots are isolated or delta type, causes this decrease in the umbral area. Another difference in Case FEM$\alpha$ from Cases FEM and FEMH is found with regard to the vertical velocity, especially in the light bridge (Fig. \ref{comp_reso}d). While the flow structure is coherent in Cases FEM and FEMH, the flow in Case FEM$\alpha$ is somewhat messy. The left (right) edge of the light bridge is mostly occupied by the downflows (upflows). However, in some segments, this rule is violated and the upflow (downflow) is observed.
\par
Apart from the light bridge, coherent downflows are also observed at the immediate edges of the umbrae (rather than at the penumbra/quiet-Sun boundaries, which is regularly seen). \cite{2011ApJ...740...15R} shows that the downflows at the umbral edges are created when the spot lacks a penumbra. Without a penumbra, the deeper layer around the umbra is exposed and the radiative energy loss becomes significant, leading to the production of such downflows. In our simulations, however, the downflows at the umbral edges are observed even in the case with prominent penumbrae (Case FEM$\alpha$). Therefore, we conjecture that the spot rotation causes these downflows. Because of the Lorentz force, the horizontal rotational flow is bent to the vertical direction and the downflow is driven there.

\subsection{Dependence of strong magnetic field}
\begin{figure}
  \centering
  \includegraphics[width=0.5\textwidth]{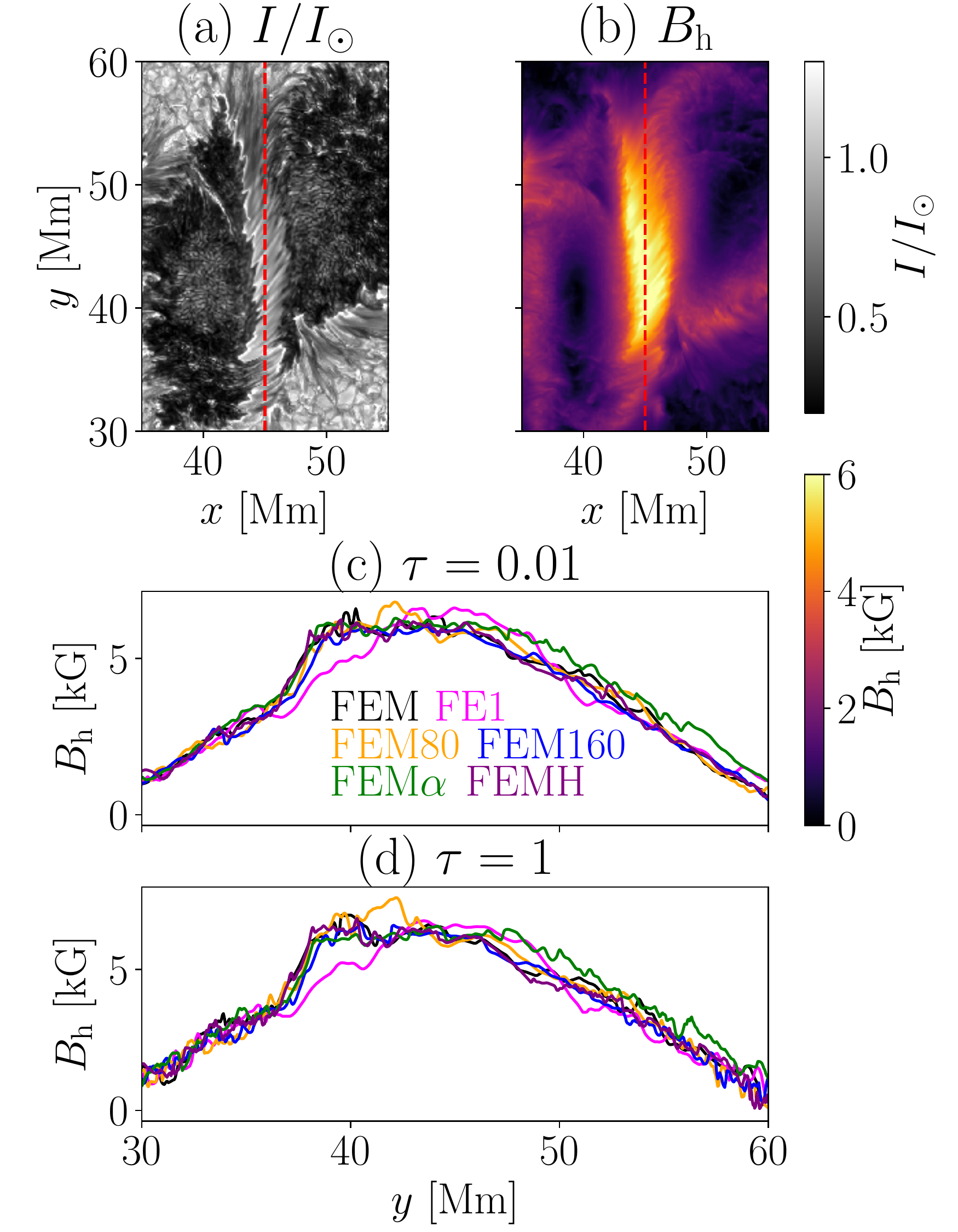}
  \caption{
    The strength of the horizontal magnetic fields in different numerical settings are compared. All data are sampled at $t=53.3~\mathrm{h}$. Panels a and b show the emergent intensity and the horizontal magnetic field strength at the $\tau=0.01$ surface for Case FEM. The cases are compared in panels c and d and show the horizontal magnetic field along the red dashed lines in panels a and b.}
  \label{horizontal_field}
\end{figure}

Fig. \ref{horizontal_field} shows the horizontal magnetic field between two sunspots at $t=53.3~\mathrm{h}$ for Case FEM. Figs. \ref{horizontal_field}a and b indicate the emergent intensity and the horizontal magnetic field strength $B_\mathrm{h}=\sqrt{B_x^2 + B_y^2}$ at $\tau=0.01$, respectively. The maximum horizontal strength in this period at $\tau=0.01$ and 1 exceeds 6600~G and 7200~G, respectively. Figs. \ref{horizontal_field}c and d present the profiles of the horizontal magnetic field strength along the red dashed line in Figs. \ref{horizontal_field}a and b, showing the results at $\tau=0.01$ and $1$, respectively. Because the density is larger on the $\tau=1$ surface than that at $\tau=0.01$, the magnetic field strength is slightly stronger on the $\tau=1$ surface. Even at $\tau=0.01$, the horizontal magnetic field strength exceeds 6000 G in all cases. Cases FEM (black), FEM80 (yellow), FEM160 (blue), and FEMH (purple) show almost the same profile of the horizontal magnetic field strength. Case FEM$\alpha$ (green) shows a slightly stronger field around $45~\mathrm{Mm} <y< 60~\mathrm{Mm}$, but the peak magnetic field around $y=40~\mathrm{Mm}$ is almost the same as for the other cases except for Case FE1. This finding makes a marked contrast with the result in \cite{2012ApJ...750...62R}. \cite{2012ApJ...750...62R} shows that a larger $\alpha$ causes a large horizontal magnetic field. While this is true in our calculations, the horizontal field of the light bridge does not depend on the top boundary condition. Case FE1 displays a slightly different magnetic field strength in some positions. In Case FE1, the calculation cannot properly resolve the fine-scale filamentary structures, potentially leading to this slight difference.
\par
\begin{figure}
  \centering
  \includegraphics[width=0.5\textwidth]{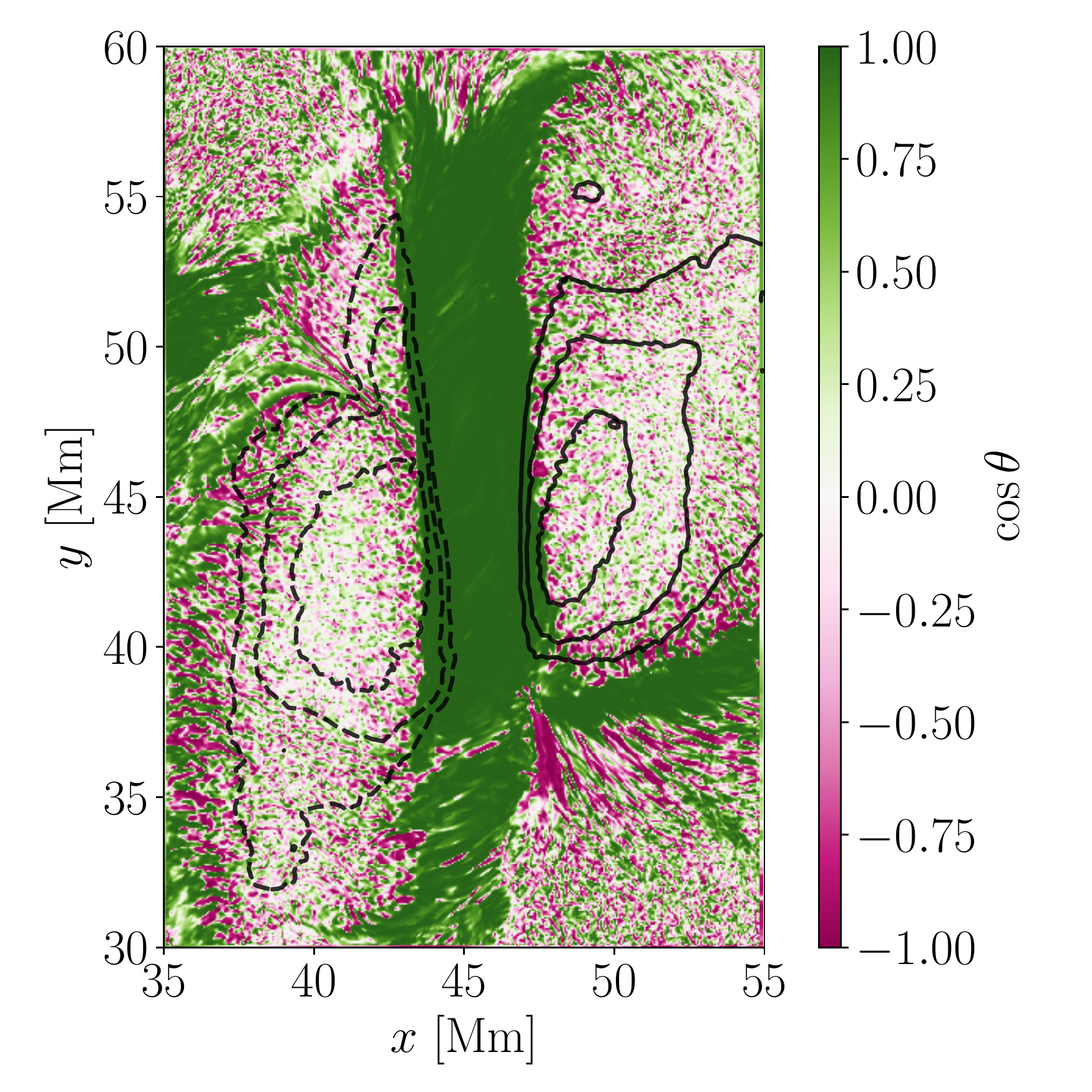}
  \caption{The cosine angle between the current and the magnetic field $(\nabla\times\bm{B})\cdot\bm{B}/|\nabla\times\bm{B}||\bm{B}|$ at $z=0$. If the value is unity, the current and the magnetic field are parallel to each other. The line contours show the vertical magnetic field $B_z$ at $z=0$. Each contour corresponds to the absolute $B_z$ of 3500, 2500, and 2000 G. The solid and dashed lines show positive and negative values, respectively.}
  \label{force_free}
\end{figure}

The consistency of the results with different numerical conditions indicates the robustness of our result, especially of the exceptionally strong magnetic fields. The fact that the different Alfven speed limits do not influence the peak strength of the horizontal magnetic field indicates that these magnetic fields are in a force-free state. Fig. \ref{force_free} shows a quantity:
\begin{align}
  \cos\theta = \frac{\left(\nabla\times\bm{B}\right)\cdot\bm{B}}{|\nabla\times\bm{B}||\bm{B}|},
\end{align}
on $z=0$. Because the Lorentz force $\bm{F}_\mathrm{L}$ is expressed as $\bm{F}_\mathrm{L}=\left(\nabla\times\bm{B}\right)\times\bm{B}/4\pi$, the force is not acting when the electric current and the magnetic field are parallel. If $\cos\theta=1$, the current and the magnetic field are parallel. Fig. \ref{force_free} shows that the force-free state is achieved within the light bridge, i.e. the strong horizontal field region.

\subsection{Amplification mechanism of the exceptionally strong horizontal magnetic field}\label{evo_mag}
In this subsection, we investigate the amplification and maintenance mechanism of the strong horizontal magnetic field in the light bridge.\par
\begin{figure}
  \centering
  \includegraphics[width=0.5\textwidth]{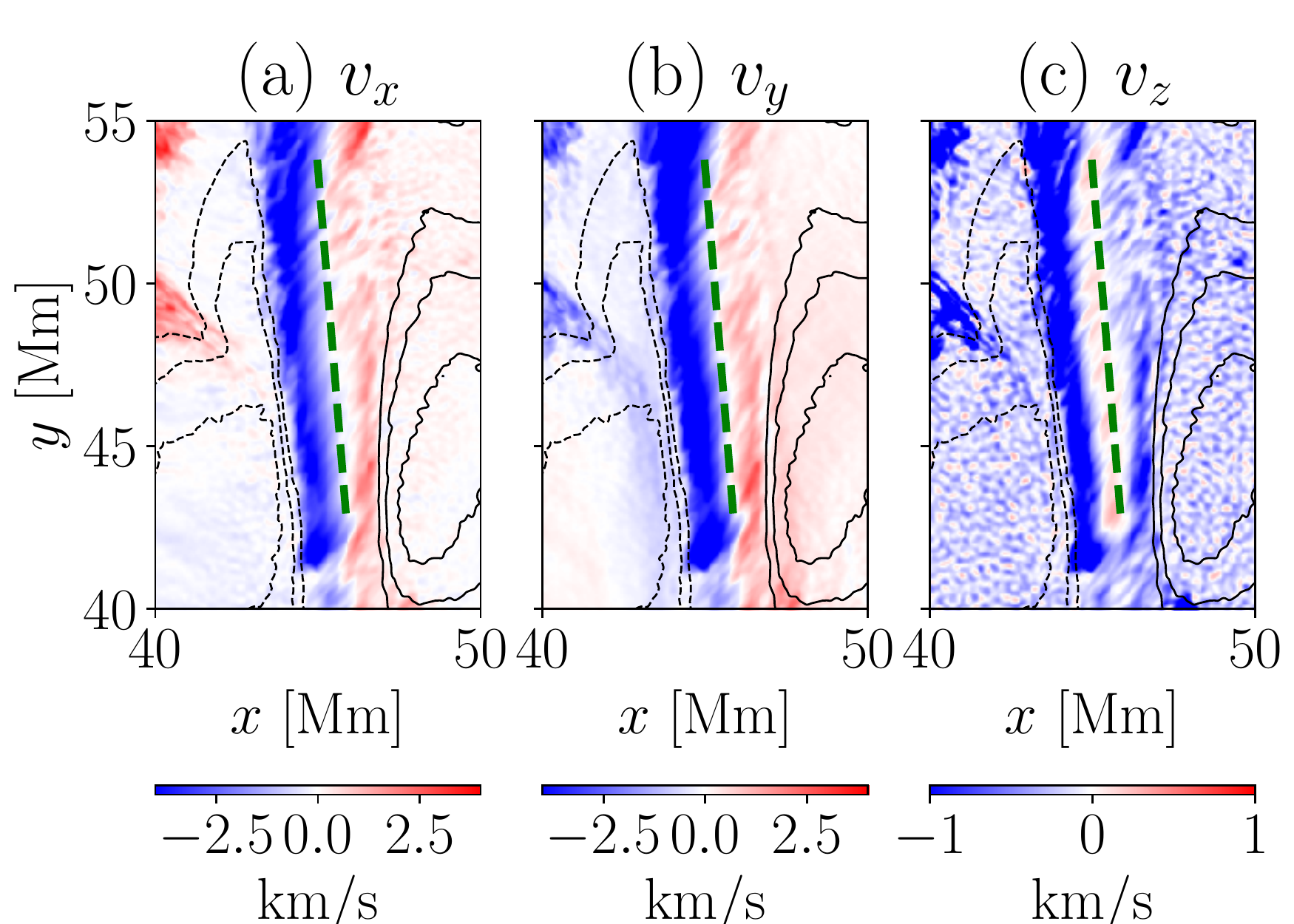}
  \caption{The flow structure at $z=0$. Panels a, b, and c show $v_x$, $v_y$, and $v_z$, respectively. The values are averaged between $t=48.3$ and $53.3$ h. Along the green dashed line in the figures, we plot the values in Fig. \ref{induction}.
  The line contours show the vertical magnetic field $B_z$ averaged over the same period. The contours correspond to 3500, 2500, and 2000 G. The solid and dashed lines indicate positive and negative values, respectively.
  }
  \label{flow_structure}
\end{figure}

Figs. \ref{flow_structure}a, b, and c show the flow structures $v_x$, $v_y$, and $v_z$ at $z=0$, respectively.
$z=0$ is at approximately the same level as the $\tau=0.01$ surface at the light bridge. Because the initial flux tube is twisted in a right-handed manner to achieve the force-free state, the sunspots in this phase are rotating clockwise by relaxing the twist. The flow structure in the light bridge reflects this rotation. At the edge of the left (right) sunspot, the horizontal flow is directed to the lower left (upper right). This means that the light bridge is filled with a divergent and sheared flow. The vertical flow in the centre of the light bridge is a weak upflow and is surrounded by the downflow. These vertical flows are also caused by the rotation.

\begin{figure}
  \centering
  \includegraphics[width=0.5\textwidth]{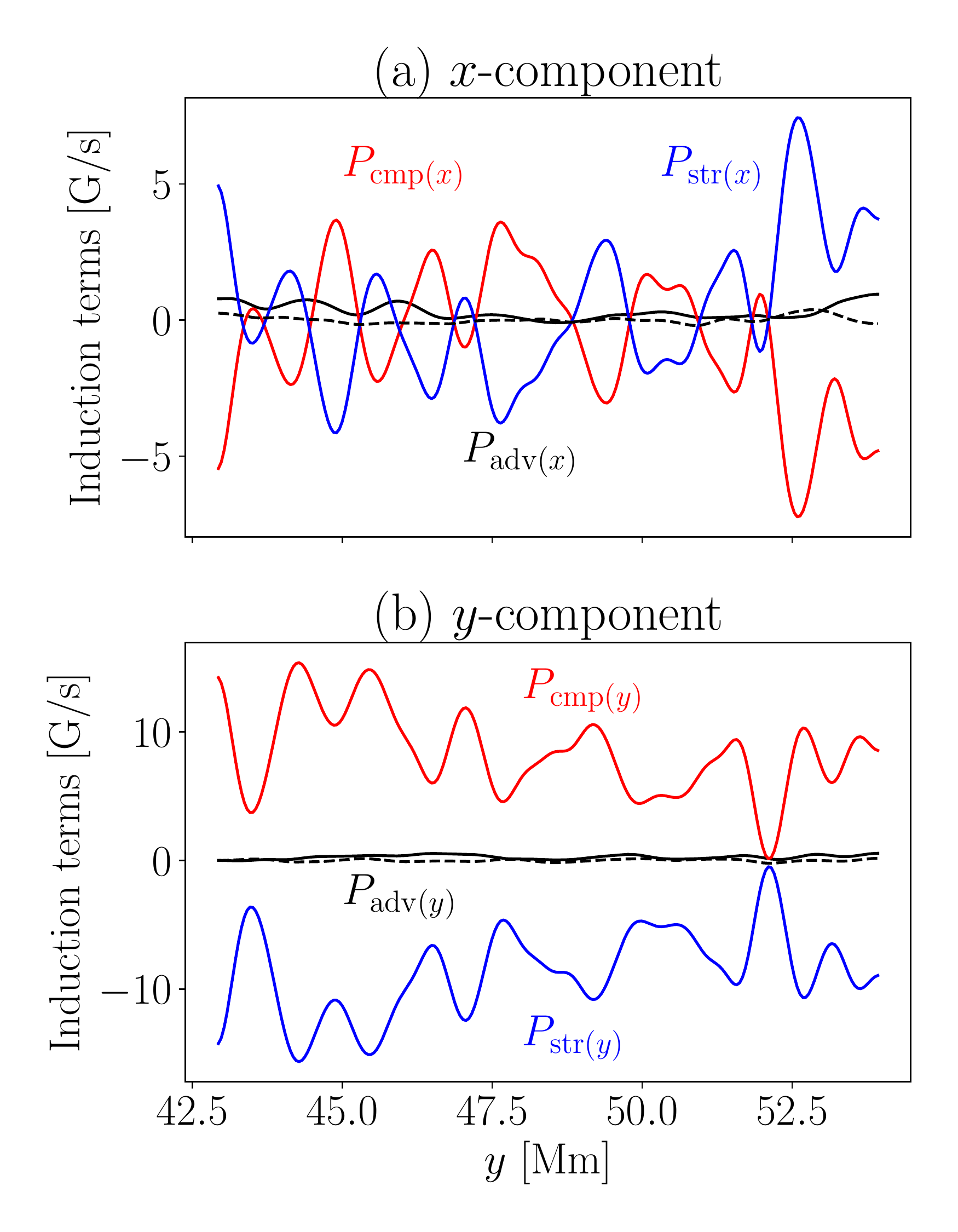}
  \caption{The terms in the induction equation along the green line in Fig. \ref{flow_structure}. 
  Panels a and b show the terms in the $x$- and $y$-component induction equation.
  The black, red, and blue lines show $P_{\mathrm{adv}(i)}$, $P_{\mathrm{cmp}(i)}$, and $P_{\mathrm{str}(i)}$, respectively. The black dashed lines show the sum of all the terms.
  }
  \label{induction}
\end{figure}

To understand the evolution of the magnetic field, we investigate the induction equation.
The induction equation is written as
\begin{align}
  \frac{\partial B_i}{\partial t} =&
  P_{\mathrm{adv}(i)} + 
  P_{\mathrm{cmp}(i)} +
  P_{\mathrm{str}(i)},
\end{align}
where $i=x,y,z$. 
The terms $P_{\mathrm{adv}(i)}$, $P_{\mathrm{cmp}(i)}$, and $P_{\mathrm{str}(i)}$ denote the advection, compression, and stretching, respectively.
These are expressed as
\begin{align}
  P_{\mathrm{adv}(i)} =& -\left(\bm{v}\cdot\nabla\right)B_i, \\
  P_{\mathrm{cmp}(i)} =& -B_i\left(\nabla\cdot\bm{v} - \frac{\partial v_i}{\partial i}\right), \\
  P_{\mathrm{str}(i)} =& \left(\bm{B}\cdot\nabla\right)v_i - B_i\frac{\partial v_i}{\partial i}.
\end{align}
In this study, we omit $B_i\partial v_i/\partial i$ from $P_{\mathrm{cmp}(i)}$ and $P_{\mathrm{str}(i)}$ because $\partial v_i/\partial i$ has no influence on the evolution of the $i$-component magnetic field ($B_i$).
We note that these terms are cancelled out when we discuss the sum of the terms.
\\
Fig. \ref{induction} shows the terms in the $x$- and $y$-component induction equation along the green dashed line in Fig. \ref{flow_structure}. The black, red, and blue lines represent $P_{\mathrm{adv}(i)}$, $P_{\mathrm{cmp}(i)}$, and $P_{\mathrm{str}(i)}$, respectively. The dashed line indicates the sum of these terms. The values are averaged over the period of $t=48.3$ to $53.3~\mathrm{h}$. While the magnetic field is evolving during this period, the sum of the terms is much smaller than the individual terms. The amplification of the magnetic field takes about 10 h (from Fig. \ref{overall}d to Fig. \ref{overall}h). Since the amplified magnetic field strength is 6000 G, the amplification rate of the magnetic field is about 0.2 $\mathrm{G\ s^{-1}}$, which is two orders of magnitude smaller than each term in the induction equations. Therefore, the best we can do is to determine the contributing term(s) for the evolution \citep[see also][]{2019A&A...631A..99S}. Regarding the $x$-component induction equation, whose individual terms are plotted in Fig. \ref{induction}a, the contributing factor is not clear. We cannot determine which term is important for the amplification and maintenance of the $x$-component field. This is discussed further in the next paragraph. Meanwhile, Fig. \ref{induction}b shows that the $y$-component magnetic field $B_y$ is amplified by the stretching. We note that the amplified $y$-component magnetic field $B_y$ at the light bridge is negative and thus the negative values for the terms correspond to the field amplification. Interestingly, the compression term weakens the horizontal field $B_y$, as this term is positive, i.e. the flow is diverging.

\par
\begin{figure}
  \centering
  \includegraphics[width=0.5\textwidth]{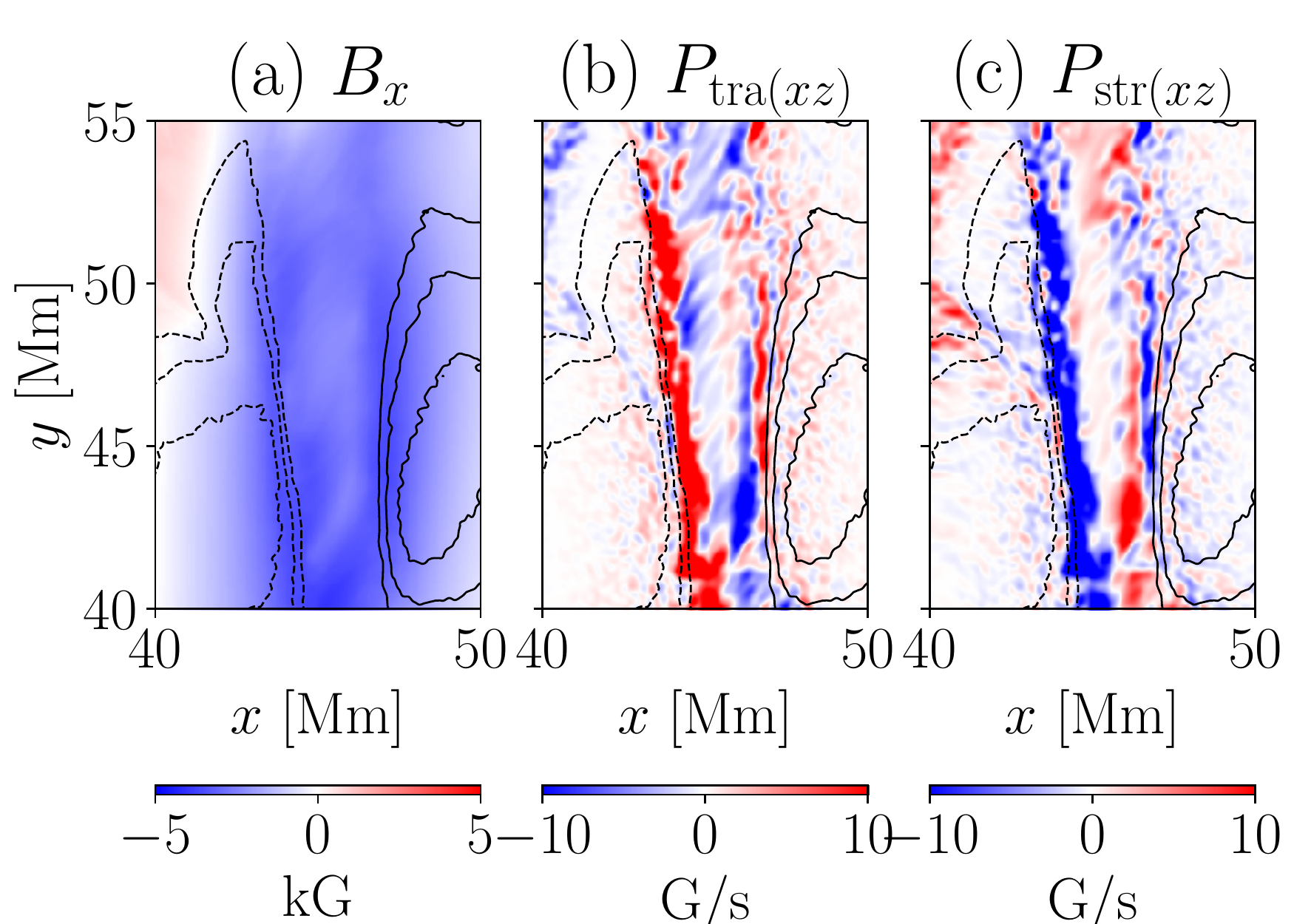}
  \caption{
    Contribution to the amplification and maintenance of $B_x$. The panels show
  (a) $B_x$, (b) $P_{\mathrm{tra}(xz)}=-\partial/\partial z\left(v_z B_z\right)$, and (c) $P_{\mathrm{str}(xz)}=B_z\partial v_x/\partial z$.
  The values are averaged between $t=48.3$ to $53.3~\mathrm{h}$. Panels b and c are smoothed with the Gaussian filter with a width of 100 km. The line contours are the same as in Fig. \ref{flow_structure}.
  }
  \label{bx_generation}
\end{figure}

We perform further analysis on the evolution of $x$-component magnetic field $B_x$ with Fig. \ref{bx_generation}. Figs. \ref{bx_generation}a, b, and c show $B_x$, $P_{\mathrm{tra}(xz)}=P_{\mathrm{adv}(xz)}+P_{\mathrm{cmp}(xz)}$, and $P_{\mathrm{str}(xz)}$, respectively, where
\begin{align}
  P_{\mathrm{tra}(xz)} =& -\frac{\partial}{\partial z}\left(v_z B_x\right),\\
  P_{\mathrm{str}(xz)} =& B_z\frac{\partial v_x}{\partial z}.
\end{align}
These terms correspond to the vertical transport of $B_x$ and the stretching of the vertical field $B_z$ in the direction of $B_x$, respectively. We find that these terms play the most important roles in enhancing the magnitude of $B_x$. In the centre of the light bridge, the vertical transport term contributes to increase $B_x$, while at the edge of the light bridge, the vertical stretching creates $B_x$. Both effects indicate that the vertical magnetic fields from the two sunspots reconnect with each other in the upper atmosphere above the light bridge (i.e. above the PIL) and create arcade-shaped field lines. Then, the amplified $B_{x}$ is transported downward and observed in the lower atmosphere at $\tau=0.01$ or $z=0$. It should be noted that the reconnection takes place probably above the top boundary because of the potential boundary condition and thus we do not directly observe the site of reconnection.

\par
\begin{figure}
  \centering
  \includegraphics[width=0.5\textwidth]{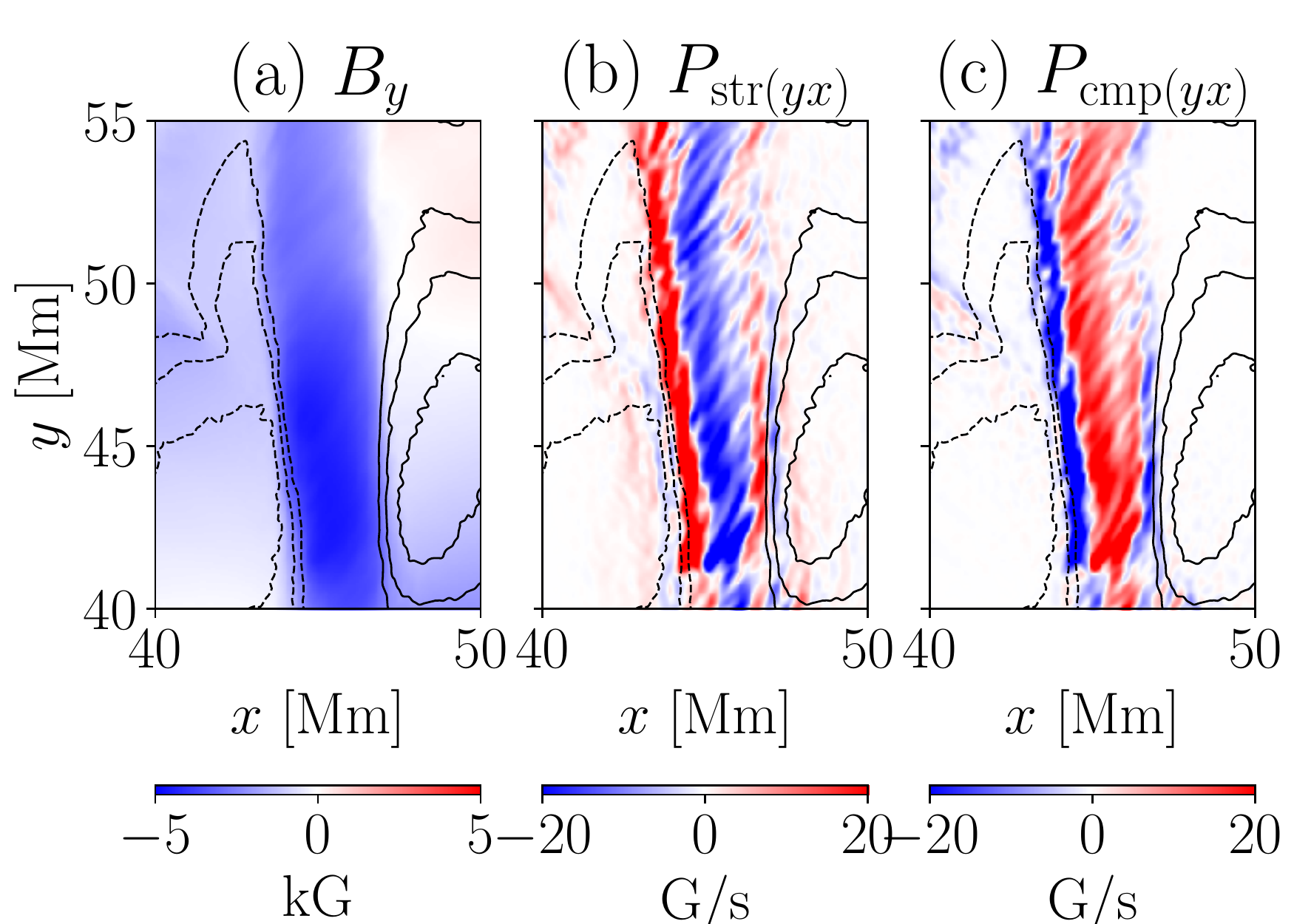}
  \caption{
  The format of this figure is the same as for Fig. \ref{bx_generation}, but the panels are (a) $B_y$, (b) $P_{\mathrm{str}(yx)}=B_x\partial v_y/\partial x$, and (c) $P_{\mathrm{cmp}(yx)}=-B_y\partial v_x/\partial x$.}
  \label{by_generation}
\end{figure}

Fig. \ref{by_generation} shows the contribution to the $y$-component induction equation. Figs. \ref{by_generation}a, b, and c describe $B_y$, $P_{\mathrm{str}(yx)}$, and $P_{\mathrm{cmp}(yx)}$, respectively, where
\begin{align}
  P_{\mathrm{str}(yx)} =& B_x\frac{\partial v_y}{\partial x}, \\
  P_{\mathrm{cmp}(yx)} =& -B_y\frac{\partial v_x}{\partial x}.
\end{align}
These terms correspond to the stretching of the $B_x$ field in the direction of $B_y$ and the compression of the $x$-component flow, respectively. Again, these terms are found to be most critical to the evolution of $B_y$. Fig. \ref{by_generation}b shows that $B_y$ in the centre of the light bridge is mainly amplified by the shear of $v_y$ (see also Fig. \ref{flow_structure}b). Because there is a divergent flow in the centre of the light bridge, the compression term decreases $B_y$. In contrast, at both edges, the compression plays a primary role in amplifying the magnetic field.

\par
\begin{figure}
  \centering
  \includegraphics[width=0.5\textwidth]{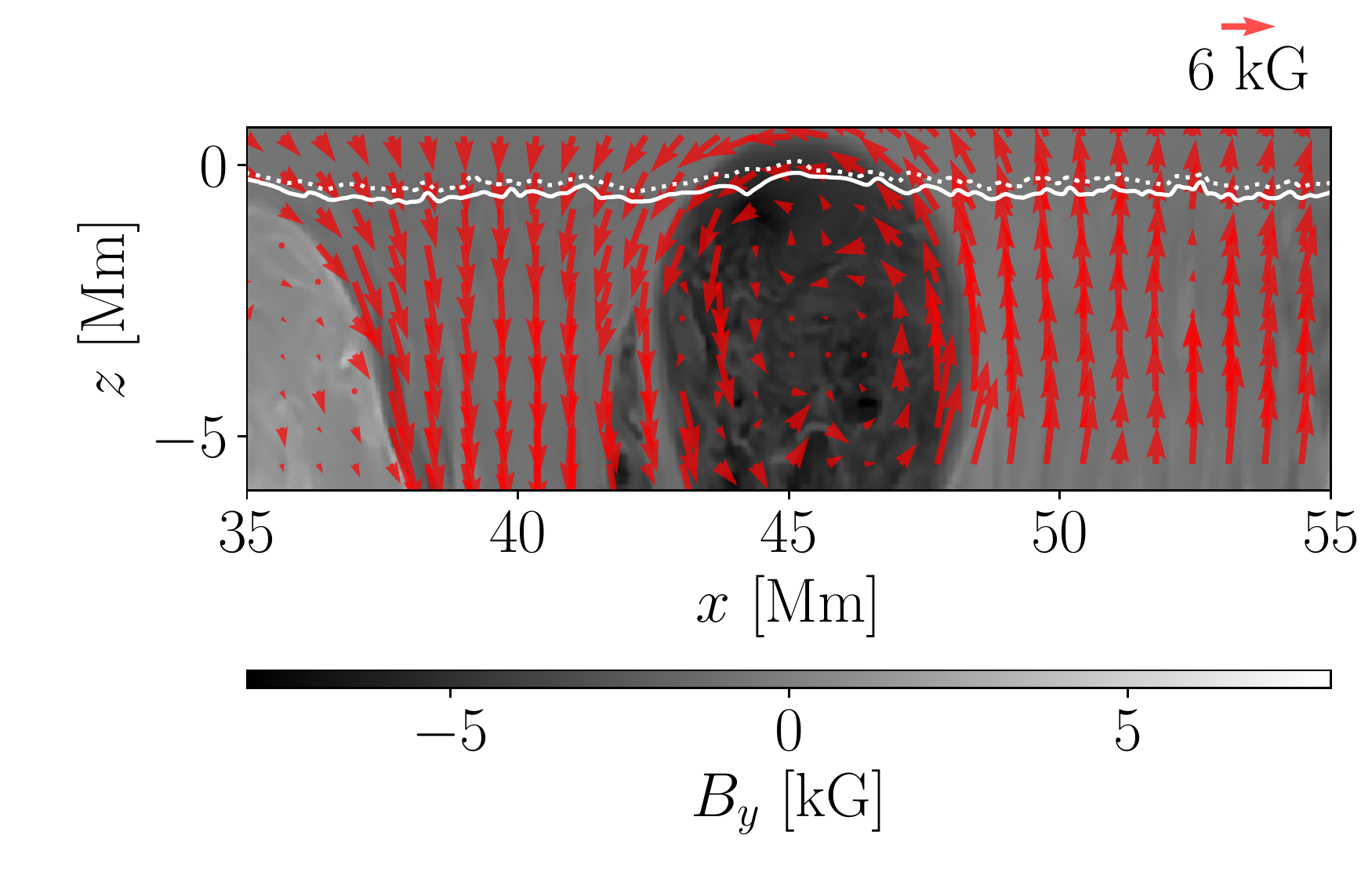}
  \caption{The vertical slice at $y=45~\mathrm{Mm}$. The red arrows show the magnetic field on the $x--z$ plane, and the colour contour represents $B_y$. The solid and dotted lines indicate the $\tau=1$ and $0.01$ surfaces, respectively.}
  \label{slice}
\end{figure}

Our finding is confirmed by the vertical slice at $y=45~\mathrm{Mm}$ (Fig. \ref{slice}). The red arrows show the magnetic field on the $x$--$z$ plane ($B_x$ and $B_z$). The greyscale represents $B_y$. The solid and dashed lines are the $\tau=1$ and $\tau=0.01$ surfaces, respectively. As discussed, the arcade-shaped magnetic field across the PIL is seen in the upper layer above $\tau=0.01$. The strong $B_y$ is created around or below the photosphere down to $-5$ Mm. As a result, a sheared arcade field is created above the PIL. The established force-free state in this region (Fig. \ref{force_free}) indicates that the magnetic pressure gradient of the axial field $B_y$ and the magnetic tension of the other components $B_x$ and $B_z$ are balanced.

\par
\begin{figure}
  \centering
  \includegraphics[width=0.5\textwidth]{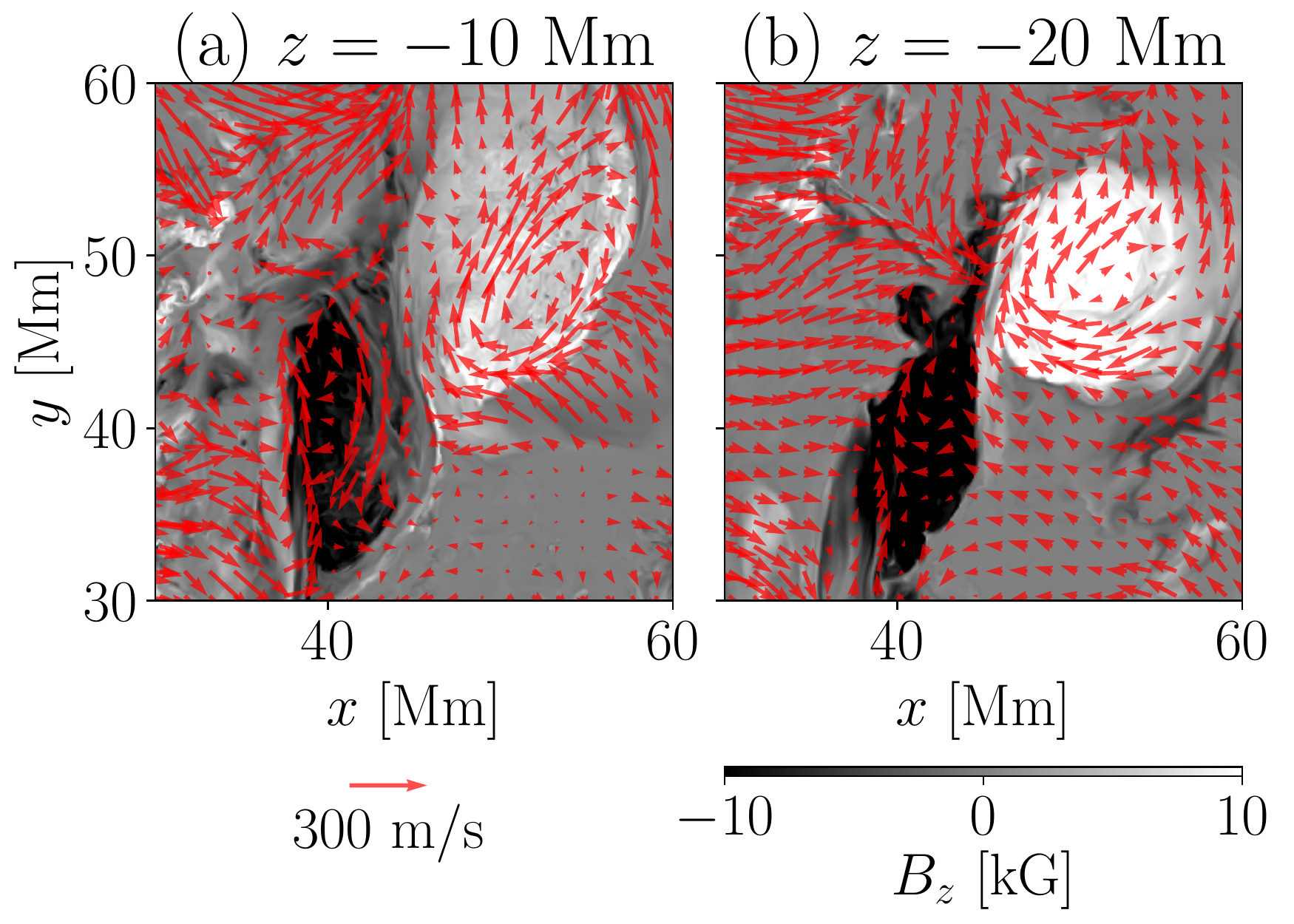}
  \caption{The flow and the magnetic field at (a) $z=-10$ and (b) $-20~\mathrm{Mm}$. The greyscale and the arrows are the vertical magnetic field $B_z$ and the horizontal flow, respectively.}
  \label{deep}
\end{figure}

Fig. \ref{deep} shows the horizontal flow ($v_x$ and $v_y$: the red arrows) and the vertical magnetic field ($B_z$: colour contour) at $z=-10~\mathrm{Mm}$ (panel a) and $z=-20~\mathrm{Mm}$ (panel b). At $z=-10~\mathrm{Mm}$, both the left and right sunspots show clockwise rotations, which is consistent with the photosphere. In the deeper layer ($z=-20~\mathrm{Mm}$), only the right sunspot appears to rotate.

\section{Summary and Discussion}
\label{Summary}

\begin{figure}
  \centering
  \includegraphics[width=0.5\textwidth]{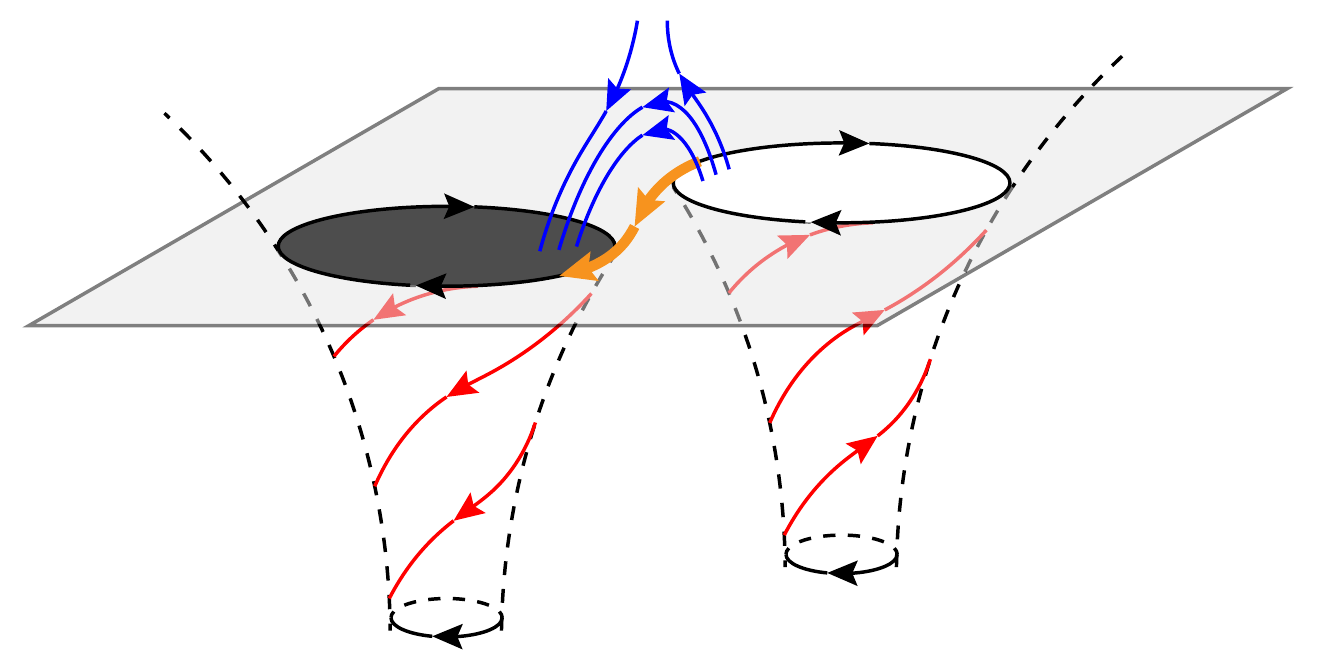}
  \caption{Summary of the generation mechanism of the strong horizontal magnetic field. The black arrows show the direction of the flows. The red arrows are the direction of the magnetic field of the initial twisted flux tube. The blue arrows indicate the magnetic field across the PIL caused by magnetic reconnection between the vertical magnetic field lines in the upper atmosphere. The orange arrows show the resulting strong magnetic field amplified by the shear of the horizontal flow.}
  \label{summary}
\end{figure}

We perform high-resolution radiative magnetohydrodynamic calculations for the formation of a strong horizontal magnetic field accompanied by delta-type spots. We cover the whole convection zone with state-of-the-art R2D2 code, and the large-scale turbulent convection generates the delta-type spot. Our main conclusions are listed as follows.
\begin{itemize}
  \item A greater than 6000~G magnetic field is reproduced at the $\tau=0.01$ surface in the light bridge between the positive and negative sunspots.
  \item The peak strength of the horizontal magnetic field does not depend on the resolution, top boundary condition, or Alfven speed limit.
  \item The amplified sheared field (i.e. the delta-spot light bridge) is almost entirely force-free because of its twisted nature.
  \item The essential amplified mechanism of the strong horizontal magnetic field is the shear motion caused by the rotation of the two sunspots.
  \item The origin of the rotation of the sunspots is the relaxation of the initial twist of the flux tube, which is rooted at a depth of at least 10 Mm.
\end{itemize}
Fig. \ref{summary} summarizes the amplification mechanism of the strong magnetic field.
\par
In this calculation, we reproduce a significantly superequipartition magnetic field. Any energy in the photosphere is not sufficient to amplify the strong magnetic field achieved in this study. Fig. \ref{deep} shows that the rotation is rooted in the deep layer, where the density is much larger than in the photosphere\footnote{The density at 10 Mm depth is more than 3600 times larger than that in the photosphere.}. The magnetic field is connected to the deep layer, and the energy there can reproduce a strong magnetic field at the photosphere. 
This is one of the key mechanisms of the superequipartition magnetic field in the photosphere.
\par
The present study shows that the contribution of the compression term is to reduce the horizontal field strength because the flow field is diverging in the light bridge because of the filamentary granular convection. This result is in stark contrast to the previous idealized delta-spot simulations without thermal convection by \cite{2017ApJ...850...39T}, in which the compression maintains the horizontal field because of the converging motions of the two spots. This demonstrates the importance of performing realistic flux emergence simulations.
\par
The essential amplification mechanism of the strong horizontal magnetic field is the stretching by the horizontal shear motion. This indicates that the resulting magnetic field should be aligned with the PIL. 
The magnetic field is, however, not aligned with it, especially in the final stage (Fig. \ref{overall}h). Because the light bridge locates between the opposite-polarity sunspots, there is plenty of magnetic flux available immediately above the PIL to amplify the magnetic field, which arches over the PIL through magnetic reconnection. If the strong horizontal field were not produced by the shear motion, the overlying field across the PIL would sink into the convection zone because of its magnetic tension. In reality, however, this downward motion is inhibited by the magnetic pressure of the horizontal field. Therefore, the overlying field across the PIL piles up on the light bridge, and the resultant magnetic field is not necessarily parallel to the PIL. It should be noted that while the flare-prolific delta-spots tend to possess highly sheared PILs \citep{2019LRSP...16....3T}, in which the filamentary convection cells are almost aligned along the PIL, some of the strongest fields have been observed in the PILs where the convection cells are less sheared and connect the neighbouring spots \citep[e.g.][]{2018RNAAS...2....8W}. Thus, the field across the PIL can be accumulated, and the resulting magnetic field does not have to be parallel to the PIL.
\par
While the generation mechanism of the exceptionally strong field is the shearing motion from the realistic simulations conducted in this study, we do not exclude other scenarios, such as the compression of umbral fields because of the surface Evershed flow, suggested by \cite{2018ApJ...852L..16O}. Yet, the key question is how to create the superequipartition field in the photosphere, and the possible answer is the magnetic linkage to the deep layer. 
\par
The origin of the sunspot rotation is the initial twist of the flux tube, which depends on our initial simulation setup. The twist begins to be relaxed as the rising starts since the gas pressure has a role in the force balance, and the relaxation process is the origin of the sunspot rotation. Also, when the flux tube reaches the photosphere, it expands due to the significant decrease of the density. This motion relaxes the twist and causes the sunspot rotation at the photosphere. The origin of the twist in the flux tube may be the deep convection and global dynamo action. We need to carry out more comprehensive calculations to address the ultimate origin of the sunspot rotation. In this study, the initial twist is determined to achieve the force-free flux tube. It is not easy to observe the twist in the deep convection zone directly. Our setup tends to show a somewhat larger rotation rate compared with the observation \citep[see TH19, ][]{2009SoPh..258..203M}. In this regard, the initial twist might be stronger than reality. Still, the top boundary condition also affects the rotation rate \citep{2010ApJ...720..233C} and we need more comprehensive investigations of the relation between the initial twist, the rotation and the top boundary condition to confirm the validity of the initial twist in our simulation.
\par
The Reynolds numbers are much smaller and the Prandtl numbers are much larger than those in the actual Sun. Our comparison between FEM and FEMH, i.e., different resolution, shows that the difference in the Reynolds numbers causes minor influence on the strong magnetic field. The higher Reynolds numbers in the actual Sun would cause more efficient small-scale dynamo; thus, the small-scale feature may be affected by the low Reynolds numbers in the simulation. The difference in the Prandtl numbers mainly causes the small-scale features \citep{2011ApJ...741...92B}.
\par
The spot rotation originated in the deep layer drives the shear motion in the photosphere in this study. However, there would be several other ways to use the energy in the deep convection zone for the generation of the strong magnetic field in the photosphere through a shearing motion. For example, the translational motion of the flux tubes and passing each other in proximity (i.e. fly-by) may cause the shearing of the spots in the photosphere, leading to the generation of a strong magnetic field in between. To understand the generation of a variety of sunspot fields in the photosphere, we need to perform a parameter study of sunspot formation simulations in the future.

\section*{Acknowledgements}
The authors would like to thank the anonymous referee for the helpful suggestions.
The results were obtained using 
Fujitsu PRIMERGY CX600M1/CX1640M1(Oakforest-PACS) at the Joint Center for Advanced High Performance Computing (JCAHPC; proposal nos. are hp190183 and hp200124) and Cray XC50 at the Center for Computational Astrophysics, National Astronomical Observatory of Japan.
This work was supported by MEXT/JSPS KAKENHI  (grant nos. JP20K14510 (PI: H. Hotta), JP15H05814 (PI: K. Ichimoto), and JP18H03723 (PI: Y. Katsukawa)), the NINS  program for cross-disciplinary study (grant nos. 01321802 and 01311904) on Turbulence, Transport, and Heating Dynamics in Laboratory and Astrophysical Plasmas: “SoLaBo-X” and MEXT as “Program for Promoting Researches on the Supercomputer Fugaku”  (Toward a unified view of the universe: from large-scale structures to planets).

\section*{Data Availability}
Data available on request.

\appendix
\section{Radiation transfer with multi-rays}
\label{multi_ray}
We explained our method for the one-ray radiation transfer in HI20. In this study, we adopt the multi-ray radiation transfer. We explain additional aspects of the radiation transfer in this appendix.\par
We solve the radiation transfer equation with the grey approximation with the Rosseland mean opacity:
\begin{align}
  \frac{\partial I}{\partial \tau} = - I + S, 
\end{align}
where $I$, $S$, and $\tau$ are the radiative intensity and $S=\sigma T^4/\pi$ is the source function and $\sigma$ is the Stefan--Boltzmann constant. As for the angular quadrature, we adopt Carlson set A4 quadrature \citep{carlson1963numerical}, where the direction-cosine $\bm{\mu}_m=\left(\mu_{x(m)},~\mu_{y(m)},~\mu_{z(m)}\right)$ = 
$\left(\sqrt{7/9},~1/3,~1/3\right)$, $\left(1/3,~\sqrt{7/9},~1/3\right)$, $\left(1/3,~1/3,~\sqrt{7/9}\right)$. The point weight is $\omega_m=1/3$ for all the $m$. This angular quadrature defines three directions per octant. Therefore, in total, we solve the 24 rays.
\par
\begin{figure}
  \centering
  \includegraphics[width=0.3\textwidth]{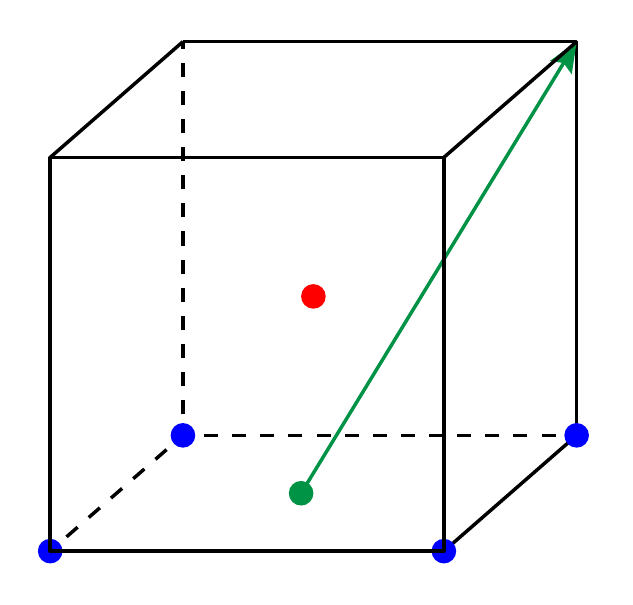}
  \caption{ Schematic of a radiation ray and the locations where the values are defined.}
  \label{rte}
\end{figure}

Fig. \ref{rte} shows the schematic of a radiation ray (green line) and the locations where the values are defined.
In the R2D2, the MHD variables are defined in the cell centre (red point).
To evaluate the variables at the cell vertex (blue points), we use the linear interpolation with eight neighbouring cell centre variables in logarithmic space as
\begin{align}
  \log Q_{i+1/2,j+1/2,k+1/2} = \frac{1}{8}
  (&
    \log Q_{i  ,j  ,k  } + 
    \log Q_{i  ,j  ,k+1} \nonumber\\
    +&\log Q_{i  ,j+1,k  } + 
    \log Q_{i  ,j+1,k+1} \nonumber\\
    +&\log Q_{i+1,j  ,k  } + 
    \log Q_{i+1,j  ,k+1} \nonumber\\
    +&\log Q_{i+1,j+1,k  } + 
    \log Q_{i+1,j+1,k+1}
  ). 
\end{align}
To evaluate the variables on a surface for the upstream of a ray (green point), again we use the linear interpolation with four cell vertex variables in the logarithmic space.\par
The length of the ray $\Delta l_m$ in a cell is calculated as
\begin{align}
  \Delta l_m = \min
  \left(
    \frac{\Delta x}{\mu_x},
    \frac{\Delta y}{\mu_y},
    \frac{\Delta z}{\mu_z}
  \right),
\end{align}
where $\Delta x,~\Delta y,$ and $~\Delta z$ are the grid spacings in $x,~y$, and $z$ directions, respectively.
The optical depth along each radiation ray $\Delta \tau_m$ is calculated as
\begin{align}
  \Delta \tau_m = \int_0^{\Delta l_m} \rho\kappa d\left(\Delta l'_m\right) \label{rte1},
\end{align}
where $\kappa$ is the opacity. Then, the intensity on the downstream value on a vertex $(I_{\mathrm{d}(m)})$ is calculated with the formal solution of the radiation transfer equation with the upstream value
$(I_{\mathrm{u}(m)})$ as:
\begin{align}
  I_{\mathrm{d}(m)} =& I_{\mathrm{u}(m)}\exp\left(-\Delta \tau_m\right) \nonumber\\
  & + \int_0^{\Delta \tau_m} S\left(\Delta \tau'_m\right)
  \exp\left(-\Delta \tau_m + \Delta \tau'_m\right)d\left(\Delta \tau'_m\right). \label{rte2}
\end{align}
For evaluation of Eqs. (\ref{rte1}) and (\ref{rte2}), we use the same scheme as HI20.\par
For evaluating the radiation heat, we need to obtain the mean intensity $J$ and the radiation flux $\bm{F}$, which are defined as:
\begin{align}
  \bm{F} =& \int_{4\pi}I\bm{\mu}d\omega = 4\pi \sum_m\omega_m \bm{\mu}_m I_m, \\
  J =& \frac{1}{4\pi}\int_{4\pi}Id\omega = \sum_m\omega_m I_m.
\end{align}
First, we evaluate the radiation flux and the mean intensity at each vertex with the presented formula. Then, we interpolate these to the cell surface and the cell centre, respectively, as:
\begin{align}
  F_{x(i+1/2,j,k)} = \frac{1}{4}
  (
     &F_{x(i+1/2,j+1/2,k+1/2)} \nonumber\\
    +&F_{x(i+1/2,j+1/2,k-1/2)} \nonumber\\
    +&F_{x(i+1/2,j-1/2,k+1/2)} \nonumber\\
    +&F_{x(i+1/2,j-1/2,k-1/2)}
  ),\\
  F_{y(i,j+1/2,k)} = \frac{1}{4}
  (
     &F_{y(i+1/2,j+1/2,k+1/2)} \nonumber\\
    +&F_{y(i+1/2,j+1/2,k-1/2)} \nonumber\\
    +&F_{y(i-1/2,j+1/2,k+1/2)} \nonumber\\
    +&F_{y(i-1/2,j+1/2,k-1/2)}
  ),\\
  F_{z(i,j,k+1/2)} = \frac{1}{4}
  (
     &F_{z(i+1/2,j+1/2,k+1/2)} \nonumber\\
    +&F_{z(i+1/2,j-1/2,k+1/2)} \nonumber\\
    +&F_{z(i-1/2,j+1/2,k+1/2)} \nonumber\\
    +&F_{z(i-1/2,j-1/2,k+1/2)}
  ),
\end{align}
and
\begin{align}
  J_{i,j,k} = \frac{1}{8}
  (
      & J_{i+1/2,j+1/2,k+1/2}
      + J_{i+1/2,j+1/2,k-1/2} \nonumber\\
      +&J_{i+1/2,j-1/2,k+1/2}
      + J_{i+1/2,j-1/2,k-1/2} \nonumber\\
      +&J_{i-1/2,j+1/2,k+1/2}
      + J_{i-1/2,j+1/2,k-1/2} \nonumber\\
      +&J_{i-1/2,j-1/2,k+1/2}
      + J_{i-1/2,j-1/2,k-1/2} 
  ).
\end{align}
Then, the radiative heating is calculated as
\begin{align}
  Q_F =& - \nabla\cdot\bm{F} \\
  Q_J =&  4\pi \kappa \rho \left(J - S\right).
\end{align}
We follow the method suggested by \cite{1999A&A...348..233B} for evaluating the radiative heating $Q_\mathrm{rad}$ as:
\begin{align}
  Q_\mathrm{rad} = Q_J\exp\left(-\frac{\tau}{-\tau_0}\right) + Q_F\left[1 - \exp\left(-\frac{\tau}{\tau_0}\right)\right],
\end{align}
where $\tau$ is the optical depth measured from the top boundary and $\tau_0=0.1$. In an optically thick layer, $J$ and $S$ take similar values and thus the accuracy of $Q_J$ decreases. The radiative flux in the optically thick layer is almost vertical, and $Q_F$ keeps the accuracy. Thus, we adopt $Q_F$ there. In contrast, the radiative flux in an optically thin layer becomes more isotropic; the orientation of the flux influences the radiative heating significantly. Therefore, in an optically thin layer, we adopt $Q_J$ instead.

\label{lastpage}
\end{document}